\documentclass[a4paper,11pt]{article}
\pdfoutput=1
\usepackage{jheppub}
\usepackage{shuffle}
\usepackage{graphicx}
\usepackage{subcaption}
\usepackage{url}
\usepackage{xcolor}
\usepackage{easy-todo}
\usepackage{here}
\usepackage{slashed}
\usepackage{booktabs,array}
\usepackage{amsmath}
\usepackage{mathtools}
\usepackage{cleveref}
\usepackage{siunitx}
\usepackage{enumitem}

\usepackage[normalem]{ulem}
\allowdisplaybreaks

\newcommand{\gram}[1]{\Delta_3^{(#1)}}

\newcommand{\vij}[1]{v_{#1}}
\newcommand{\mHsq}{q^2}
\newcommand{\mTsq}{m_t^2}
\newcommand{\cayley}[1]{\mathcal{C}_{#1}}

\newcommand{\amflow}{\textsc{AMFlow}}
\newcommand{\rambo}{\textsc{Rambo}}

\newcommand{\sqrtalg}[1]{\sqrt[\uproot{5}\scriptscriptstyle\flat]{#1}}

\newcommand{\sqrtC}{\sqrt{\mathcal{C}_1}}
\newcommand{\sqrtG}{\sqrt{\Delta_3^{(1)}}}
\newcommand{\sqrtGG}{\sqrt{\Delta_3^{(2)}}}
\newcommand{\sqrtGfv}{\sqrt{\Delta_3^{(5)}}}

\title{Pentagon Functions for Two-Loop Leading-Colour $pp\to t\bar{t}H$ Amplitudes
with a Light-Quark Loop}

\author[1]{Simone Devoto,}\emailAdd{simone.devoto@ugent.be}
\author[2]{Manfred Kraus,}\emailAdd{mkraus@fisica.unam.mx}
\author[1]{Ben Page}\emailAdd{ben.page@ugent.be}

\affiliation[1]{Department of Physics and Astronomy, Ghent University, 9000
Ghent, Belgium}
\affiliation[2]{Departamento de F\'{i}sica Te\'{o}rica, Instituto de
F\'{i}sica, \\ Universidad Nacional Aut\'{o}noma de M\'{e}xico, Cd. de
M\'{e}xico C.P. 04510, M\'{e}xico}

\abstract{
We present a basis of special functions and an efficient numerical implementation for the planar two-loop master integrals that enter the leading-colour amplitudes for $pp\to t\bar{t}H$ with a closed light-quark loop.
We provide analytic results for the special functions at transcendental weight one and two in terms of logarithms and dilogarithms, while higher weights are evaluated as one-fold integrals.
We focus on the nested roots that arise, studying their analytic structure on the physical region and constructing numerically stable representations of letters and $\mathrm{d}\log$ forms that involve them. 
We find that a number of $\mathrm{d}\log$ forms have an antiderivative that cannot be represented by a single logarithm over the physical region and tackle them numerically with adaptive Chebyshev approximation.
We implement the functions in \texttt{C++} as an extension of the \texttt{PentagonFunctions++} library. We study the performance of our implementation over physical phase-space points and find that the evaluation is accurate and fast enough for phenomenological application. This provides an important building block for NNLO studies of $t\bar{t}H$ production at the LHC.
}

\preprint{
\begin{flushright}
  \includegraphics[width=2cm]{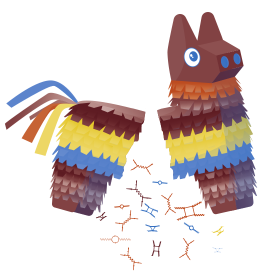}
\end{flushright}
}

\begin{document}
\maketitle
%--------------------------------------------------------------------------------
%--------------------------------------------------------------------------------
\section{Introduction}
\label{sec:introduction}
%--------------------------------------------------------------------------------
One of the main goals of the high-luminosity physics programme of the LHC is the study of the properties of the Higgs boson.
With the increased dataset that it will provide, we are expecting first meaningful constraints on the Higgs boson self-coupling, together with an increased precision of the measurements of its coupling with the other Standard Model particles.
The top-quark Yukawa coupling is the largest one of them, and can be directly accessed at the LHC through the associated production of a top-quark pair and a Higgs boson, $pp\to t\bar tH$.
Since its observation at the LHC~\cite{CMS:2018uxb,ATLAS:2018mme}, measurements of this process have constrained deviations from the Standard-Model Yukawa coupling at the level of ten percent~\cite{ATLAS:2022vkf,CMS:2022dwd},
and by the end of the high-luminosity run they are expected to reach the level of a few percent, at which point theoretical uncertainties will dominate~\cite{Cepeda:2019klc,ATLAS:2025eii}.
This makes it essential to include next-to-next-to-leading-order (NNLO) QCD predictions in phenomenological studies.
Predictions at this level of precision are by now available, where the so far missing two-loop amplitudes have been approximated in different kinematic limits, with the resulting uncertainty on the predictions estimated at the percent level.
This was done first for the inclusive cross section~\cite{Catani:2021cbl,
  Catani:2022mfv}, where the two-loop amplitude was approximated by an expansion
in the soft-Higgs limit, then for differential
distributions~\cite{Devoto:2024nhl}, where a high-energy expansion in the small
top-quark mass limit was also used.
Recently, such fixed-order results have also been matched to resummation~\cite{Balsach:2025jcw} and parton showers~\cite{Biello:2026nhj}.
Further developments in the calculation of the amplitudes include an
approximate computation in the boosted limit~\cite{Wang:2024pmv}, and their
exact numerical evaluation in the quark-initiated $N_f$
part~\cite{Agarwal:2024jyq} and in the leading-colour approximation for the $q\bar q$ channel~\cite{Chen:2026bxb}.
Removing the need for these approximations at NNLO calls for two-loop master integrals that can be evaluated quickly and reliably everywhere in physical phase space. 
In this paper we provide such an implementation for the planar master integrals entering the leading-colour two-loop amplitudes with a closed light-quark loop.

These master integrals were first presented in ref.~\cite{FebresCordero:2023pww}, representing the first two-loop 
five-point integral computation in a problem with seven kinematic scales and massive internal lines.
There, a basis of master integrals was constructed that satisfies a system of differential equations in $\epsilon$-factorised form. 
The master integrals were then evaluated numerically via generalised series expansions~\cite{Moriello:2019yhu} as implemented in \texttt{DiffExp}~\cite{Hidding:2020ytt}, with boundary values from the auxiliary mass flow method~\cite{Liu:2017jxz,Liu:2018dmc,Liu:2022chg,Huang:2026rjb}.
This yielded high-precision benchmarks, but evaluating the integrals at the large number of phase-space points required by an NNLO calculation remained challenging.
An approach to overcome this challenge is the construction of so-called ``pentagon  functions''~\cite{Caron-Huot:2014lda,Gehrmann:2018yef,Chicherin:2020oor}. 
One identifies, at each transcendental weight, a minimal set of algebraically independent special functions in terms of which all master integrals can be expressed. 
Functions of weights one and two are written in terms of logarithms and dilogarithms, while functions of weights three and four are represented as one-fold integrals along a straight-line path in phase space~\cite{Caron-Huot:2014lda}. 
Implemented in the \texttt{PentagonFunctions++}
library~\cite{Chicherin:2020oor,Chicherin:2021dyp,Abreu:2023rco}, this approach
provides the two-loop special functions underlying many of the available NNLO
QCD predictions for $2\to3$ processes (see ref.~\cite{Huss:2025nlt} for a recent
overview).
Alternative strategies for evaluating multi-scale integrals include generalised series expansions~\cite{Moriello:2019yhu,Armadillo:2022ugh,Hidding:2020ytt,Lee:2017qql,Liu:2022chg,Huang:2026rjb}, Chebyshev approximations~\cite{Abreu:2026vxw,Liu:2026cpf}, numerical differential equation solving~\cite{PetitRosas:2025xhm,Badger:2025ljy,Czakon:2026tog} and finite-element approaches~\cite{Chen:2026pyw}.
In this work, we follow the one-fold integral approach, which exploits directly the compact $\epsilon$-factorised differential equation and the analytic form of the functions at low weight.

Applying the pentagon-function approach to the $t\bar tH$ integrals raises new issues, which stem from the presence of nested square roots in the alphabet (a feature that has been observed since then in other complicated two-loop integrals~\cite{Badger:2024fgb,Gotzfried:2026def,Becchetti:2025oyb,Becchetti:2025qlu,Bera:2026qxx}).
The nested roots are non-analytic on a hyperplane in the physical phase space,
even though this hyperplane is not a Landau singularity.
We avoid the difficulties associated with crossing this hyperplane by placing the boundary point
of our integration paths on the hyperplane.
Furthermore, straightforward representations of the letters that involve nested
roots, and of their $\mathrm{d}\log$ forms, develop spurious singularities,
which we remove by suitable rewriting.
Finally, the $\mathrm{d}\log$ forms of some letters have antiderivatives that
cannot be written as a single logarithm over the whole phase space.
In order to avoid partitioning phase space, we construct these antiderivatives
numerically along the integration path with adaptive Chebyshev
approximations~\cite{Abreu:2026vxw}.

We implement the evaluation of our set of special functions in \texttt{C++}, as an extension of \texttt{PentagonFunctions++}~\cite{Chicherin:2020oor,Chicherin:2021dyp,Abreu:2023rco}, which can be downloaded from the public repository:

{
\medskip
\centering \href{https://gitlab.com/pentagon-functions/PentagonFunctions-cpp}{\texttt{https://gitlab.com/pentagon-functions/PentagonFunctions-cpp}}\par
\medskip
}

\noindent Comparing evaluations in double and quadruple precision over a sample of physical phase-space points, we find that the numerical accuracy is under excellent control, and the average evaluation of the full set of functions takes around half a second,
making the functions suitable for phenomenological applications. 
As the same integrals appear in two-loop amplitudes for processes such as $pp\to t\bar tZ$, our results are also relevant beyond $t\bar tH$ production.

Our work is organised as follows. In \cref{sec:kinematics} we set up the kinematics and the square roots of the alphabet. In \cref{sec:DE} we recall the integral families and their differential equations. In \cref{sec:basis} we analyse the nested square root on physical phase space and its consequences for the integration paths, and construct the basis of pentagon functions weight by weight, giving analytic expressions at weights one and two and one-fold integral representations at weights three and four. In \cref{sec:numerics} we describe the numerical implementation, including the treatment of the nested-root letters and the Chebyshev approximation of $\mathrm{d}\log$ form antiderivatives, and assess its precision and performance. We conclude in \cref{sec:conclusions}.

%--------------------------------------------------------------------------------
\section{Scattering Kinematics and Notation}
\label{sec:kinematics}
%--------------------------------------------------------------------------------
In this work, we consider the production of a top-quark pair $t\bar t$ in association with a Higgs boson $H$ at hadron colliders. Specifically, we work in the kinematic channel
\begin{align}
q_1(p_4) q_2(p_5) \to t(-p_1) H(-p_2)\bar t(-p_3) \;,
\label{eq:Channel}
\end{align}
where $q_i$ labels a generic initial-state massless parton. In particular, we work with on-shell top quarks and the on-shell conditions for the momenta take the form
\begin{align}
    p_1^2=p_3^2=m_t^2\hspace{1cm} p_2^2=q^2\hspace{1cm} p_4^2=p_5^2=0\,.
\end{align}
We keep the virtuality $q^2$ of $p_2$ generic, so that the results also apply to other massive final-state bosons, such as the $Z$ boson in on-shell $t\bar tZ$ production.
By convention, we take all external momenta $p_i$ to be incoming, so that momentum conservation reads
\begin{equation}
    \sum_{i=1}^5 p_i=0\,.
\end{equation}
The scalar Feynman integrals that we consider are functions of the Mandelstam invariants
\begin{align}
    \vec s = \{v_{12},v_{23},v_{34},v_{45},v_{15},m_t^2,q^2\}\;,
\end{align}
where we define
\begin{align}
v_{ij}=2p_i\cdot p_j\;.
\end{align}
All other invariants $v_{ij}$, such as $v_{13}$, $v_{14}$, $v_{24}$, $v_{25}$ and $v_{35}$ appearing below, are linear combinations of those in $\vec s$.

In the phase-space channel defined by \cref{eq:Channel}, the kinematic invariants respect the following set of inequalities:
\begin{align}
  \label{eqn:phasespace_bnd}
  & m_t^2 > 0 \ , 
  && q^2 > 0\ ,  
  && \vij{12}\ge 2\; m_t\; q\ ,
  \\\nonumber
  & \vij{23} \ge 2\; m_t\; q\ ,  
  && \vij{34} \le 0 \ , 
  && \vij{15} \le 0\ ,  
  \\\nonumber
  & \vij{45} \ge (2 m_t + q)^2 \ , 
  && \det G(p_i,p_j,p_k) \ge 0\ , 
  && \det G(p_1, p_2, p_3, p_4) \le 0\ ,
\end{align}
where $G(k_1,\dots, k_n)_{ij} = k_i\cdot k_j$ is the Gram matrix.
Beyond these constraints, we will also impose the additional constraint that the
final-state off-shellness is less than 4 times the square of the top mass, i.e.,
\begin{equation}
\label{eq:C1pos}
    4m_t^2>q^2\;.
\end{equation}
Importantly, \cref{eq:C1pos} is fulfilled by on-shell kinematics and for most phenomenological applications for $t\overline{t}H$ production, as well as for on-shell $t\overline{t}Z$ production.

The Feynman integrals under consideration have a number of algebraic branch points. Many of them are familiar from one-loop calculations. A number can be identified as the zeros of the following Gram determinants
\begin{align}
\gram{1} &= -4\det G(p_1,p_2) = \vij{12}^2 - 4\mTsq\mHsq\;, \\
 \gram{2} &= -4\det G(p_2,p_3) = \vij{23}^2 - 4\mTsq\mHsq\;, \\
 \gram{3} &= -4\det G(p_1,p_2+p_3) = (\mHsq+\vij{23}-\vij{45})^2 - 4\mTsq\vij{45}\;,\\ 
 \gram{4} &= -4\det G(p_1+p_2,p_3) = (\mHsq+\vij{12}-\vij{45})^2 - 4\mTsq\vij{45}\;, \\
 \gram{5} &= -4\det G(p_2,p_3+p_4) = (\mHsq+\vij{34}-\vij{15})^2 - 4\mHsq(\mTsq+\vij{34})\;, \\
 \Delta_5\ \!\ &= 16\det G(p_1,p_2,p_3,p_4).
\end{align}
Further algebraic branch points occur at zeros of the so-called ``Cayley determinants''
\begin{align}
 \cayley{1} &= \mHsq(\mHsq-4\mTsq)\;, \label{eq:c1} \\
 \cayley{2} &= \Big[(\mHsq+\vij{12})(\mHsq+\vij{23})-\mHsq\vij{45}\Big]\Big[(\mHsq+\vij{12})(\mHsq+\vij{23})-(\mHsq-4\mTsq)\vij{45}\Big] \label{eq:c2} \;.
\end{align}
The kinematic constraint that we imposed in \cref{eq:C1pos} has been chosen such that we can assume that we are always on one side of the branch point at $ \cayley{1} = 0$. In particular, $\mathcal{C}_1<0$, so $\sqrt{\mathcal{C}_1}$ is purely imaginary throughout the region we consider.
A number of the algebraic branch points of the Feynman integrals do not arise at one loop, denoted as $r_i$ in ref.~\cite{FebresCordero:2023pww}. We refer to ref.~\cite{FebresCordero:2023pww} for their explicit expressions.
Beyond this, in ref.~\cite{FebresCordero:2023pww}, in order to rotate to a basis of pure integrals, it was necessary to introduce a square root of
a quantity which already contained a root, $\sqrt{{N_\pm}}$, defined through
\begin{align}
N_\pm &= q^2\left(N_b \pm \sqrt{N_b^2 - N_c}\right), 
\label{eq:NpmDefinition}
\\
N_b &= q^2\left[(v_{14}+v_{15})^2 + (v_{34}+v_{35})^2\right] - 2m_t^2\,(v_{24}+v_{25})^2, \\
N_c &= \mathcal{C}_1\,(v_{12}-v_{23})^2\,(v_{12} + v_{23} + 2 v_{13} + 4m_t^2)^2.
\end{align}
As discussed in ref.~\cite{FebresCordero:2023pww}, the square roots of $ N_+$ and $N_-$ are not algebraically independent, but are related through their product, $N_+N_-=(q^2)^2N_c$ (see \cref{eq:NpNmProduct}), where $N_c$ is $\mathcal{C}_1$ times a perfect square. In order to simplify the algebraic branch structure of the pure basis, this led to the definition of an algebraic square root $\sqrtalg{N_-}$ as
\begin{align}
\label{eq:RootNmDefinition}
    \sqrtalg{ N_-} \equiv \sqrt{\cayley{1}}\, \frac{\mHsq(\vij{12}-\vij{23})(
    v_{12} + v_{23} + 2 v_{13} + 4m_t^2
    )}{\sqrt{ N_+}}\;.
\end{align}             
We stress that this is a definition of $\sqrtalg{N_-}$ that is distinct from the principal-branch one: it ties the branch-cut structure of $\sqrtalg{N_-}$ to that of $\sqrt{N_+}$.
In contrast, $\sqrt{N_+}$ is evaluated using the standard principal branch prescription.
As we will see in \cref{sec:nested_root}, these nested roots have a highly
non-trivial structure on physical phase space.

%--------------------------------------------------------------------------------
\section{Integral Families and Differential Equations}
\label{sec:DE}
%--------------------------------------------------------------------------------
\subsection{Integral Families}
%--------------------------------------------------------------------------------
In ref.~\cite{FebresCordero:2023pww}, it was shown that the leading-colour
fermionic contribution to the two-loop $t\bar t H$ amplitude can be expressed in
terms of a basis of independent master integrals that is closed under the
map
\begin{equation}
  \mathcal{Z} \quad : \quad p_1 \leftrightarrow p_3, \quad p_4 \leftrightarrow p_5\;.
  \label{eq:Zmap}
\end{equation}
While the one-loop master integrals can all be expressed in terms of the pentagon integral family ($T_0$), the two-loop master integrals can be expressed in terms of two integral families: the penta-box ($T_1$) and the penta-bubble ($\tilde T_2$) integral families.
We label the complete set of master integrals as $I_i$, where they are listed according to the ordering associated to each topology: $T_0$, $T_1$, $\tilde T_2$, $\mathcal Z(\tilde T_2)$.
Within a topology, the master integrals are ordered according to the conventions fixed in~\cite{FebresCordero:2023pww}, to which we refer the reader for their complete list.

\begin{figure}[t]
\begin{center}
 \includegraphics[scale=0.8]{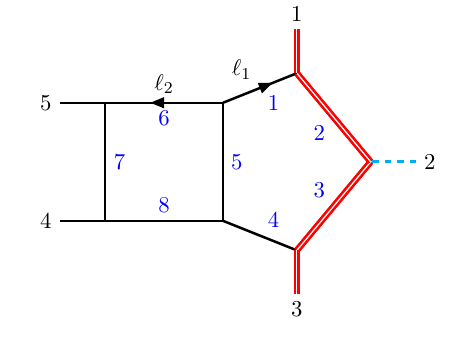}
 \hspace{5mm}
 \includegraphics[scale=0.65]{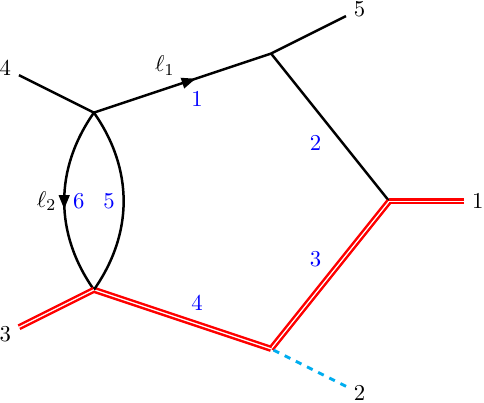}
\end{center}
 \caption{The topologies of the two integral families entering our computation: the penta-box $T_1$ on the left and the penta-bubble ($\tilde T_2$) on the right. (Diagrams reproduced from ref.~\cite{FebresCordero:2023pww}.)}
 \label{fig:topo}
\end{figure}

The penta-box integral family, labelled with $T_1$, refers to the topology of the left diagram in \cref{fig:topo}.
It is defined by 8 inverse propagators, with three additional inverse propagators introduced to complete the integral family, making it possible to express any scalar product between the loop momenta $\ell_1$, $\ell_2$ and the external momenta in terms of inverse propagators.
A generic element of the integral family $T_1$ is then identified by the vector of exponents of the propagators, $\vec \nu$, and can be written as:
\begin{equation}
 T_1[\vec{\nu}] = (e^{\gamma_E} \mu^2)^{2\epsilon}\int \frac{\mathrm{d}^d\ell_1}{i\pi^{d/2}}\frac{\mathrm{d}^d\ell_2}{i\pi^{d/2}}
 \frac{\rho_9^{-\nu_9}\rho_{10}^{-\nu_{10}}\rho_{11}^{-\nu_{11}}}
 {\rho_1^{\nu_1}\rho_2^{\nu_2}\rho_3^{\nu_3}\rho_4^{\nu_4}\rho_5^{\nu_5}\rho_6^{\nu_6}
 \rho_7^{\nu_7}\rho_8^{\nu_8}}\;,
 \label{eq:family}
\end{equation}
 where $d=4-2\epsilon$, $\mu$ is the dimensional-regularisation scale, and propagators $\rho_i$ carry the Feynman prescription $\rho_i\to\rho_i+i0$. They are defined as
\begin{align}
 &\rho_1 = \ell_1^2\;, &&\rho_2 = (\ell_1+p_1)^2 - \mTsq\;, &\rho_3 &= (\ell_1+p_{12})^2 - \mTsq\;,\nonumber \\
 &\rho_4 = (\ell_1+p_{123})^2\;, &&\rho_5 = (\ell_1+\ell_2)^2\;, &\rho_6 &=\ell_2^2\;, \label{eq:T1props} \\
 &\rho_7 = (\ell_2+p_5)^2\;, &&\rho_8 = (\ell_2+p_{45})^2\;,  &\rho_9 &= (\ell_1-p_5)^2\;, \nonumber\\
 &\rho_{10} = (\ell_2-p_{12})^2 - \mTsq\;, 
 &&\rho_{11} = (\ell_2-p_1)^2 - \mTsq\;. \nonumber
\end{align}
We note that the integral family $T_1$ is invariant under the action of the transformation $\mathcal Z$ of \cref{eq:Zmap}.
The dimension of the vector space spanned by the integral family $T_1$, and thus
the number of master integrals with this topological structure, is 111.

The penta-bubble integral family refers instead to the topology on the right of
\cref{fig:topo}.
We label it as $\tilde T_2$ and we write a generic element of this integral
family as:
\begin{equation}
 \tilde T_2[\vec{\nu}] = (e^{\gamma_E} \mu^2)^{2\epsilon} \int \frac{\mathrm{d}^d\ell_1}{i\pi^{d/2}}\frac{\mathrm{d}^d\ell_2}{i\pi^{d/2}}
 \frac{\rho_7^{-\nu_7}\rho_{8}^{-\nu_{8}}\rho_{9}^{-\nu_{9}}\rho_{10}^{-\nu_{10}}\rho_{11}^{-\nu_{11}}}
 {\rho_1^{\nu_1}\rho_2^{\nu_2}\rho_3^{\nu_3}\rho_4^{\nu_4}\rho_5^{\nu_5}\rho_6^{\nu_6}}\;.
 \label{eq:T2tfamily}
\end{equation}
In this case, there are 6 inverse propagators directly coming from the
topological structure of the diagram, and we introduce 5 additional ones to
complete the integral family.
We define them as 
\begin{align}
 &\rho_1 = \ell_1^2\;, &&\rho_2 = (\ell_1+p_5)^2\;, &&\rho_3 = (\ell_1+p_{15})^2 - \mTsq\;,\nonumber \\
 &\rho_4 = (\ell_1+p_{125})^2 - \mTsq\;, &&\rho_5 = (\ell_1+\ell_2-p_4)^2\;, &&\rho_6 =\ell_2^2\;,
\label{eq:T2tprops} \\
 &\rho_7 = (\ell_1-p_3)^2\;, &&\rho_{8} = (\ell_2-p_1)^2\;,  &&\rho_{9} = (\ell_2-p_2)^2\;, \nonumber \\
 &\rho_{10} = (\ell_2-p_3)^2\;, &&\rho_{11} = (\ell_2-p_4)^2\;.\nonumber
\end{align}
The dimension of the vector space spanned by $\tilde T_2$ is 19.
Nevertheless, of the 19 resulting master integrals, 11 are shared with the basis of $T_1$, leaving us with only 8 new master integrals.
In contrast with the case of $T_1$, we observe that $\tilde T_2$ is not invariant under the mapping given by the transformation $\mathcal Z$.
As a consequence, we need to consider also the vector space spanned by $\mathcal Z (\tilde T_2)$ which thus adds 8 new master integrals to our analysis.

When computing the finite remainder for the two-loop light-quark-loop amplitudes, it is
also necessary to have results for the associated one-loop Feynman integrals,
which we also compute in this work.
We therefore also introduce the integral family $T_0$ defined through
\begin{equation}
 T_0[\vec{\nu}] = (e^{\gamma_E} \mu^2)^{\epsilon} \int \frac{\mathrm{d}^d\ell_1}{i\pi^{d/2}}
 \frac{1}
 {\rho_1^{\nu_1}\rho_2^{\nu_2}\rho_3^{\nu_3}\rho_4^{\nu_4}\rho_5^{\nu_5}}\;,
 \label{eq:T0family}
\end{equation}
with inverse propagators defined as:
\begin{align}
 &\rho_1 = \ell_1^2\;, &&\rho_2 = (\ell_1+p_1)^2 - \mTsq\;, &&\rho_3
=(\ell_1+p_{12})^2 - \mTsq\;,\nonumber \\
 &\rho_4 = (\ell_1+p_{123})^2\;, &&\rho_5 = (\ell_1-p_5)^2\;\ .
&&\label{eq:T0props} 
\end{align}
This adds 18 master integrals to our counting.

%--------------------------------------------------------------------------------
\subsection{Differential Equations and Independent Functions}
%--------------------------------------------------------------------------------
For the leading-colour light-quark-loop contribution to $t\bar tH$ production,
ref.~\cite{FebresCordero:2023pww} identified the basis of master integrals $\vec
I$ that brings the corresponding differential equation into $\epsilon$-factorised
form:
\begin{align}
    \mathrm{d} I_i = \epsilon \, M_{ijk} \; \mathrm{d}\log(W_j) \;I_k\;.  \label{eq:canonical}
\end{align}
A basis with such a property is known as a ``canonical basis''.
We call the possible arguments $W_j$ of the  $\mathrm{d}\log$ forms appearing in
the system of differential equations in
\cref{eq:canonical} ``letters'', and
their collection $\mathcal A$ the ``alphabet''. We refer the reader to
ref.~\cite{FebresCordero:2023pww} for a complete description of the alphabet,
which consists of 152 letters.\footnote{In ref.~\cite{FebresCordero:2023pww},
  differential forms $\omega_{130}, \ldots, \omega_{133}$ were not written as
  the $\mathrm{d}\log$ of an algebraic letter. Using the \texttt{Trager} package
  of ref.~\cite{PagePoegelTrager}, we found such a representation as $\omega_{i}
  = \frac{1}{2}\mathrm{d}\log(W_i)$, which we use throughout. The expressions
  for these four $W_i$ are provided in ancillary files described in
  \cref{sec:ancillaryFiles}.
}

The $\epsilon$-factorised form of the differential equation is a crucial feature
that facilitates its solution. In particular, let us consider the $\epsilon$
expansion of the master integrals,
\begin{align}
I_i (\vec{s}, \epsilon) = \sum_{n=0}^\infty \epsilon^n I^{(n)}_{i}(\vec{s}) \;,
\end{align}
which starts at $n=0$ due to the basis choice of ref.~\cite{FebresCordero:2023pww}.
Notably, the components $I^{(0)}_i$ are rational constants.
If we insert this form into \cref{eq:canonical} we obtain a tower of
differential equations at each order in $\epsilon$ as
\begin{align}
\label{eq:deI}
\mathrm{d} I_{i}^{(n)} = M_{ijk}\;\mathrm{d}\log(W_j)\;I_{k}^{(n-1)} \;.
\end{align}
Integrating along a path $\gamma$ gives
\begin{align}
  {I}^{\;(n)}_i(\vec{s}_1\!\;) = {I}^{\;(n)}_i(\vec{s}_0) + M_{ijk}\int_\gamma \mathrm{d} \log(W_j)\,{I}_k^{\;(n-1)}\,,
\label{eq:Iiter}
\end{align}
where $\gamma$ is a path over phase space that connects the boundary point $\vec
s_0$ to the desired evaluation point $\vec s_1$.
Given a boundary value $I^{(n)}_i(\vec s_0)$ (that can be obtained numerically,
e.g.\ with \amflow{}~\cite{Liu:2022chg,Huang:2026rjb}), \cref{eq:Iiter} formally allows one
to compute the $\epsilon$ expansion of the master integrals.

In order to evaluate the master integrals in an efficient way, we will express them in terms of a smaller subset of independent special functions of fixed transcendental weight.
We will refer to these functions as \emph{pentagon functions} and label them as $f_i^{(n)}$, where the index $n$ identifies the transcendental weight of the functions, while $i$ enumerates them. 
To identify them, we proceed weight by weight.
At each weight $n$, we first identify a basis $g^{(n)}$ of the vector space spanned by the master integrals.
We choose the elements of $g^{(n)}$ as a linearly independent subset of the $\epsilon$-expanded elements $I^{(n)}_i$ of the canonical basis, as is standard (see e.g.\ refs.~\cite{Badger:2021nhg,Gehrmann:2024tds,Abreu:2023rco}).
The components of the master integrals are related to such basis by the relation
\begin{align}
\label{eq:pentagon}
I_i^{(n)}=A^{(n)}_{ij}g_j^{(n)}\;,
\end{align}
where $A^{(n)}$ is a matrix of rational numbers.
By inserting \cref{eq:pentagon} into \cref{eq:deI}, we get the system of
differential equations satisfied by the linearly independent master integrals
\begin{equation}
\label{eq:pentagonDE}
    \mathrm{d} g_i^{(n)}
   =c^{(n)}_{ijk}\;\mathrm{d}\log(W_j)\;g_k^{(n-1)}
    \;,
\end{equation}
where the $c^{(n)}_{ijk}$ are rational numbers, which implies the recursive formula, along the same path as in \cref{eq:Iiter}:
\begin{align}
    \label{eq:pentagonIT}
     g_i^{(n)}(\vec{s}_1) = g_i^{(n)}(\vec{s}_0) + c_{ijk}^{(n)} \int_\gamma \mathrm{d} \log( W_j )\; g_k^{(n-1)}\;.
\end{align}
Not all elements of $g^{(n)}$ are new, however. In practice, there exist combinations of the $g_i^{(n)}$ that can be written as
polynomials of lower-weight functions and $\zeta$ values.
We call the elements that are irreducible modulo products of lower-weight
functions the pentagon functions $f_j^{(n)}$, identifying them using the procedure
of ref.~\cite{Chicherin:2021dyp}.
As such, all $g_i^{(n)}$ can be written as a polynomial of the $f_j^{(n')}$ and $\zeta(n')$ where $n' \le n$.
We note that, conventionally, we choose the zeta value of weight~$n$ to be a pentagon function, i.e. it is one of the $f_j^{(n)}$.
In summary, $I^{(n)}_i$ can be expressed as linear combinations of pentagon functions of
the corresponding weight and products of lower-weight pentagon functions. 
The corresponding relations can be found in the ancillary files attached to this
publication.

%--------------------------------------------------------------------------------
\section{Pentagon Function Construction}
\label{sec:basis}
%--------------------------------------------------------------------------------

In this section we carry out the construction of \cref{sec:DE} weight by weight,
up to weight~4.
The numbers of linearly independent and irreducible functions at each weight are
given in \cref{tab:funcspace}.
At weights 1 and 2, we construct explicit
representations for the pentagon functions in terms of logarithms and
dilogarithms.
At weights 3 and 4, we instead use one-fold
integral representations that follow from \cref{eq:pentagonIT}.
Many of the steps in our construction are informed by the analytic structure of
the differential equation.
We thus undertake an analysis of the alphabet on physical phase space, which
guides the construction at each weight as well as a study of the analytic
structure of the nested square root which has strong implications for our
choice of integration path at weights 3 and 4.

\begin{table}[t]
\centering
\begin{tabular}{ccc}
 \toprule
 $n$ & $\# g^{(n)}$ & $\# f^{(n)}$ \\
 \midrule
 1 & 7 & 7 \\
 2 & 33 & 17 \\
 3 & 103 & 70 \\
 4 & 141 & 120 \\
 \bottomrule
\end{tabular}
\caption{Number of elements of the basis $g^{(n)}$ of the space spanned by the $I_i^{(n)}$, \cref{eq:pentagon}, and of the pentagon functions $f^{(n)}$, the elements irreducible modulo products of lower-weight functions, at each weight $n$. For $n\geq2$, both counts include the zeta value $\zeta(n)$.}
\label{tab:funcspace}
\end{table}

\subsection{Behaviour of the Alphabet on Physical Phase Space}
\label{sec:alphabet_phys}
Up to weight 4, the functions constituting our basis can be expressed in terms
of the 143 ``relevant'' letters introduced in ref.~\cite{FebresCordero:2023pww}
($9$ further letters contribute only at higher weights).
The letters can be categorised depending on their properties under Galois transformations: we define as \emph{even letters} the ones that have trivial Galois transformations, while we define as \emph{odd letters} the ones picking up a sign under such transformations.
Since these letters will enter the arguments of our special functions, it is
particularly important to know their behaviour when evaluated over physical
phase-space points, as it determines whether logarithmic branch cuts involving
the letters can be kept away from the physical region.
In practice, proving sign properties for every letter is an involved task. We
therefore sidestep it by studying the letters numerically over physical phase
space. The following observations are therefore empirical.

We find that the letters can be divided in two sets:
\begin{itemize}
    \item $\mathcal A_\mathbb{R}$: letters with fixed sign, either positive or negative. All the even letters fall into this category, and part of the odd letters. 
    \item $\mathcal A_\mathbb{C}$: letters that are a pure phase. All the odd letters that do not have a definite sign fall in this category, and thus lie on the unit circle on the complex plane.
\end{itemize}
Importantly, we find that none of the purely real letters can vanish over phase space, a feature which is greatly helpful for numerical stability when solving the differential equation.
For example, for $W\in\mathcal A_\mathbb{R}$ the sign in $\log(\pm W)$ can be chosen once and for all such that the logarithm is real-analytic over the whole physical region.
To understand the branch-cut structure of special functions having pure-phase letters as an argument, it is important to know which of the elements of $\mathcal A_\mathbb{C}$ can take any possible phase.
We find that for only 6 elements of $\mathcal{A}_\mathbb{C}$ does the image of the physical region cover the whole unit circle. These are given by
\begin{align}\label{eq:circles}
   \mathcal A_\mathbb{C}^\circ= \{
    W_{60}, W_{75}, W_{118}, W_{128}, W_{132}, W_{133}
    \}\,.
\end{align}
For the other elements of $\mathcal A_\mathbb{C}$, the image of the physical
region is a proper arc of the unit circle, which does not contain one of $W = 1$
or $W = -1$.
This property implies, for example, that one can choose the sign of $\log(\mp
W)$ such that the branch cut is not crossed by any path in the physical region.
In contrast, for the letters in $\mathcal A_\mathbb{C}^\circ$ no such choice
exists: whatever the position of the branch cut, some physical paths cross it.
These properties are used throughout the construction. At weight 2, they
determine which combinations of dilogarithms are continuous over phase space,
while at weight 4, they inform the technique used to construct the
anti-derivatives of $\mathrm{d}\log(W)$.
A full categorisation of the letters can be found in appendix~\ref{sec:letters_prop}.

%--------------------------------------------------------------------------------
\subsection{Weights 1 and 2: Special Function Representations}
\label{sec:weight1And2}
%--------------------------------------------------------------------------------
We now discuss our construction of the pentagon functions at low weights. For weights 1 
and 2 specifically, as is standard, we find representations in terms of well-known special 
functions: logarithms and dilogarithms. Explicit expressions can be found in ancillary files that we 
detail in \cref{sec:ancillaryFiles}.

At weight 1, the $I_i^{(1)}$ span a 7-dimensional space, and all 7 of the
$g_i^{(1)}$ are irreducible and hence can be identified as pentagon functions.
The analytic expression for the weight 1 pentagon functions can be
straightforwardly obtained by integrating \cref{eq:pentagonIT} and obtaining
logarithms of the letters up to a possible transcendental constant.
The transcendental constants can be fixed, for example, by comparing the
evaluation of the analytic expressions at different phase space points with
numerical ones obtained with public tools such as
\amflow{}~\cite{Liu:2022chg,Huang:2026rjb}.
In this way, we define a basis of weight 1 pentagon functions as
\begin{equation}
\begin{aligned}
f_1^{(1)} &= -\log\left(\frac{m_t^2}{\mu^2} \right)\,, \qquad
& f_2^{(1)} &= \log\left(\frac{q^2+v_{12}}{m_t^2}\right)-i\pi\,,
\\
f_3^{(1)} &= \log\left(\frac{q^2+\sqrt{\cayley{1}}}{q^2-\sqrt{\cayley{1}}}\right)-i\pi\,, \qquad
& f_4^{(1)} &= -\log\left(\frac{v_{45}}{\mu^2} \right)+i\pi\,,
\\
f_5^{(1)} &= \log\left(\frac{q^2+v_{23}}{m_t^2}\right)-i\pi\,, \qquad
& f_6^{(1)} &= \log\left(-\frac{v_{15}}{m_t^2}\right)\,,
\\
f_7^{(1)} &= \log\left(-\frac{v_{34}}{m_t^2}\right)\,.
\end{aligned}
\label{eq:weight1Defs}
\end{equation}
We note that the scale $\mu$ enters only $f_1^{(1)}$ and $f_4^{(1)}$. It arises from expanding the overall factor $(\mu^2)^{L\epsilon}$ of an $L$-loop integral, which multiplies lower orders of the $\epsilon$ expansion.
Naturally, a basis of simpler combinations of logarithms is available and the
particular ones in \cref{eq:weight1Defs} arise because we choose our basis to
match Feynman integral components.
Importantly, the arguments of the logarithms appearing in \cref{eq:weight1Defs}
never cross the branch cut of the logarithm when considered over physical phase space.

At weight 2, the corresponding construction is more intricate.
We again begin by counting: the weight 2 master integrals can be written in terms of 17 irreducible weight 2 pentagon functions and products of the weight 1 pentagon functions and can be entirely expressed in terms of polylogarithms.
We construct analytic expressions for the weight 2 functions in terms of
logarithms and dilogarithms making use of
symbol-based
techniques~\cite{Goncharov:2005sla,Goncharov:2010jf,Duhr:2011zq,Duhr:2012fh}.
Specifically, we follow the procedure of ref.~\cite{Duhr:2011zq}, which has
frequently been applied to Feynman integrals (see e.g.\
refs.~\cite{Bonciani:2016qxi,Bonciani:2019jyb,vonManteuffel:2017myy,Heller:2019gkq}).

One of the key steps in the procedure of ref.~\cite{Duhr:2011zq} is to identify
a set of dilogarithms that can be used to express the weight 2 functions, modulo
products of logarithms.
To this end, one starts by identifying the possible arguments such that the
symbol of the corresponding dilogarithm can be expressed in terms of letters of
the alphabet.
In order to avoid terms that live outside of the alphabet, the approach of
ref.~\cite{Duhr:2011zq} searches for arguments $\prod_i W_i^{\alpha_i}$ with
exponents $\vec\alpha$ that satisfy
\begin{align}
\label{eq:wdilog}
    1 - c \prod_i W_i^{\alpha_i} = \hat{c} \prod_i W_i^{\beta_i}\;.
\end{align}
for some other exponent vector $\vec\beta$ and constants $c$ and $\hat{c}$.
Once such a set is identified, an analytic representation of the weight 2
function can be easily constructed using linear algebra at the symbol level,
supplemented by numerical evaluations to fix possible constant terms.

\begin{figure}[t]
\begin{center}
 \includegraphics[scale=1]{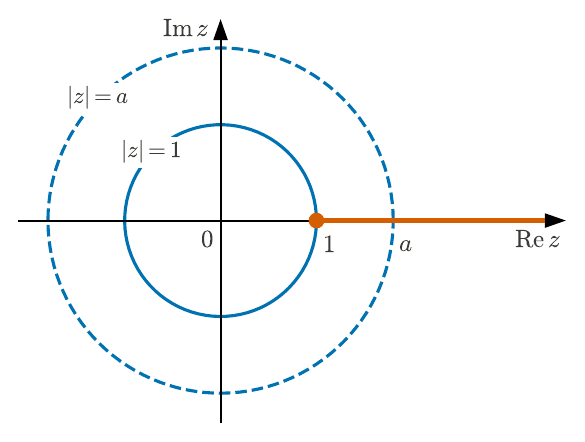}
\end{center}
\caption{Branch cut of $\mathrm{Li}_2(z)$ (in orange) together with the two kinds of complex arguments $z$ of the dilogarithms appearing in the pentagon functions (in blue)\: pure phases, which stay on the unit circle, and arguments $z=a\,e^{i\varphi}$, whose modulus $a$ also varies over phase space (dashed, shown for one value of $a$).}
\label{fig:branch}
\end{figure}

In practice, we find that the set of dilogarithms with arguments simultaneously
satisfying~\cref{eq:wdilog} and of sufficiently low degree to be computationally
tractable is insufficient to completely span the space of product-projected
weight 2 functions modulo products of logarithms and constants.
To remedy this, we introduce ``auxiliary letters'' built by decomposing the odd
letters into numerator and denominator factors. Such auxiliary letters allow the
procedure of ref.~\cite{Duhr:2011zq} to succeed at much lower degree.
Specifically, we restricted the search to candidates whose numerator and
denominator have mass-squared dimension at most $4$.
As an example of an auxiliary letter, we split the odd letter $W_7$ as 
\begin{equation}
    W_7=\frac{W_{7,A}}{W_{7,B}}\;, \qquad
    W_{7,A}=  q^2+\sqrtC\,, \quad  W_{7,B}=  q^2-\sqrtC\,.
\end{equation}
A complete list of all the auxiliary letters and their definition can be found in appendix~\ref{sec:aux_letters}.
Importantly, the auxiliary letters do not enlarge the space of $\mathrm{d}\log$ forms spanned by the alphabet.
This follows from the well-known observation that the Galois invariant product of
numerator and denominator factors gives products of even letters.
For instance, the relation $W_{7,A}W_{7,B} = 4W_6 W_8$ implies that 
\begin{align}
    \mathrm{d}\log W_{7,A} = \frac12\,\mathrm{d}\log(W_6W_8) + \frac12\,\mathrm{d}\log W_7\, .
  \label{eq:auxiliaryDlogRelationExample}
\end{align}
In the same way, up to the choice of branches, every auxiliary letter factorises
over the even and odd letters of the original alphabet.

In our application of the procedure of ref.~\cite{Duhr:2011zq}, we put
particular care into the selection of a basis of a set of dilogarithms such that
we could represent each weight 2 function with a single representation that
covers all of physical phase space.
This implies selecting arguments for the $\text{Li}_2$ functions that manifest
the analytic properties of the pentagon functions. Specifically, the weight 2
pentagon functions are continuous over phase space, so we looked for ways to
build $\text{Li}_2(z)$ (or simple combinations thereof) that also satisfy this
property.
Since every auxiliary letter factorises over the letters of the alphabet, every dilogarithm argument, including those containing auxiliary letters, can be written, up to the choice of branches, as
\begin{equation}
\label{eq:zgeneral}
z = c \prod_i |W_{\mathbb R, i}|^{\zeta_i}\prod_j W_{\mathbb C,j}^{\eta_j}\;,\quad
\zeta_i,\eta_j\in\mathbb Q\;,\quad
W_{\mathbb R, i}\in\mathcal A_{\mathbb R}\;,\quad
W_{\mathbb C, j}\in\mathcal A_{\mathbb C}\;.
\end{equation}
Here $c$ is a constant, which also absorbs the constant phases arising from
fractional powers of negative letters and from the choice of branches.
For the arguments we use, no letter of $\mathcal A_\mathbb{C}^\circ$ appears.
The fractional powers of the pure-phase letters are therefore continuous over
phase space.
Moreover, as we are working at weight 2, the collection of complex factors
that arise is greatly restricted. Naturally, complex values arise due to the presence of
square roots in the letters. In practice, the set of square roots consists of
\begin{align}
\label{eq:rootsW2}
    \sqrtC\,,\quad \sqrt{\mathcal C_2}\,,\quad \sqrt{\Delta_3^{(k)}}\,,\ k=1,\ldots,5\,.
\end{align}
Out of this set, only $\sqrt{\mathcal C_1}$ is imaginary in the physical region,
while the other six roots are real.
Importantly, this means that in the following we can identify complex
conjugation with the Galois transformation $\sqrtC \to -\sqrtC$, which helps to
organise our analysis.

In practice, we find that we can classify the different arguments $z$ into three categories
\begin{itemize}
\item $\vec \eta=0$. In this case $z$ is $c$ times a positive quantity, so its
  phase is that of $c$ and is fixed over phase space. For the arguments that we use,
  $c$ is real, so $z$ is real with a fixed sign.
  This allows us to introduce dilogarithms in our basis of weight 2 functions that, over phase-space, never cross the branch cut starting at $z = 1$.
In particular, for $z<0$, we make use of $\text{Li}_2(z)$ while, for $z>0$, we
make use of $\text{Li}_2(1-z)$.
\item $\vec \zeta=0$ and $|c|=1$. In this case, $z$ is a pure phase, a complex number with
modulus $1$: over phase space, $\text{Li}_2(z)$ is thus continuous, since the
unit circle meets the branch cut $[1,\infty)$ only at $z=1$, where $\text{Li}_2$
is continuous (see \cref{fig:branch}).
However, while all the complex letters that arise in $z$
are odd under $\sqrtC\to-\sqrtC$, $\text{Li}_2(z)$ in general is not.
For weight two functions that are odd under the associated Galois
transformation, it is natural to choose a basis for the $\text{Li}_2$ piece
which is also odd.
To this end, we explicitly make use of the function
\begin{align}
    \text{D}_\circ(z)=\text{Li}_2(z)-\text{Li}_2(\bar{z})\;,
\end{align}
which is manifestly odd, since $\sqrtC\to-\sqrtC$ maps $z$ to $\bar z$.
On the unit circle, $z=e^{i\theta}$, one has $\text{D}_\circ(z) =
2i\,\text{Im}\,\text{Li}_2(z)$ from which we can see that $\text{D}_\circ$ is
purely imaginary and continuous on the whole unit circle.

\item $\vec \zeta\neq0$, $\vec\eta\neq0$. In this case, $z$
      has the form $z=a\,e^{i\varphi}$.
      Unlike the previous two cases, the form of $z$ does not prevent it from
      reaching the branch cut (see \cref{fig:branch}). Rather than limiting
      ourselves to arguments where this does not happen, we use
\begin{align}
    \text{E}_2(z)=\text{Li}_2(z)+\text{Li}_2(\bar z)\;\,
\end{align}
    which is continuous for any $z$, since the discontinuity of $\text{Li}_2$ across
    $[1,\infty)$ is purely imaginary. Moreover, $\text{E}_2$ is even under
    $\sqrtC\to-\sqrtC$, which maps $z$ to $\bar z$ and it therefore is a
    natural building block of Galois even functions.
\end{itemize}
In this way, each weight 2 pentagon function is given by a single expression in terms of $\text{Li}_2$, $\text{D}_\circ$ and $\text{E}_2$, products of logarithms and constants, which is continuous over the whole physical region and so needs no analytic continuation. These expressions are provided in the ancillary files (see \cref{sec:ancillaryFiles}) and serve as input to the one-fold integral representations at weights 3 and 4.

%--------------------------------------------------------------------------------
\subsection[The Nested Square Root and Paths in Physical Phase Space]{The Nested Square Root and Paths in Physical Phase Space}
\label{sec:nested_root}
%--------------------------------------------------------------------------------

Beyond weight 2, our approach to solving the differential equations rests upon
representing the solution as a one-fold integration along a path, as in
\cref{eq:Iiter}. At weights 3 and above, the iterated integration representation
in \cref{eq:Iiter} involves letters $W_i$ that depend on the nested square roots $\sqrt{N_\pm}$.
These nested square roots turn out to have a non-trivial analytic structure that
influences our choice of integration path. In this subsection, we therefore analyse
this structure.
We first discuss how the physical phase space constraints fix the signs of the
arguments of the nested roots, and then identify where the nested roots fail to
be analytic, which restricts the integration paths we may use.

To begin, we observe that the signs of $N_\pm$ are fixed over physical phase
space. From \cref{eq:NpmDefinition}, we have
\begin{equation}
    N_+N_-=(q^2)^2N_c\,,
    \label{eq:NpNmProduct}
\end{equation}
which is non-positive throughout physical phase space as $\mathcal{C}_1<0$ by
\cref{eq:C1pos} and the remaining factors of $N_c$ are squares.
Thus, for $N_c \ne 0$, $N_+$ and $N_-$ have opposite signs.
Moreover, the non-positivity of $N_c$ implies that $\sqrt{N_b^2-N_c}\ge|N_b|$
and we can therefore see that on physical phase space
\begin{equation}
    N_+\geq0,
    \qquad
    N_-\leq0.
\end{equation}
In particular, $\sqrt{N_+}$ is real throughout physical phase space, while both $\sqrtalg{N_-}$ and $\sqrt{N_-}$ are purely imaginary.

Let us now determine where the inner root vanishes, which is where it could give
rise to an algebraic branch point. Using $\mathcal{C}_1<0$, its argument can be
written as a sum of two manifestly non-negative quantities,
\begin{equation}
    N_b^2-N_c
    =
    N_b^2
    +
    |\mathcal{C}_1|(v_{12}-v_{23})^2
    (v_{12} + v_{23} + 2 v_{13} + 4m_t^2)^2\,.
    \label{eq:InnerPositivity}
\end{equation}
Importantly, if the right-hand-side of \cref{eq:InnerPositivity} is to vanish,
each of these non-negative quantities must vanish separately.
Given our physical phase space assumptions, the factors $|\mathcal{C}_1|$ and
$v_{12}+v_{23}+2v_{13}+4m_t^2$ of the second term are strictly positive on
physical phase space. (For the latter, note that
$v_{13}=(p_1+p_3)^2-2m_t^2\ge2m_t^2$).
Consequently, within physical phase space we have that
\begin{equation}
    N_b^2-N_c=0 
    \qquad \Longleftrightarrow \qquad
    v_{12} = v_{23},
    \qquad
    N_b = 0.
\end{equation}
We will denote this surface as $\mathcal{S}_0$.
Notably, we see that $\mathcal{S}_0$ is a codimension two surface, as it
involves two constraints. We therefore do not need to design any
special procedure to avoid this branch point surface, since generic paths avoid it.

\begin{figure}[t]
    \centering
    \begin{subfigure}[c]{0.48\textwidth}
        \centering
        \includegraphics[width=\textwidth]{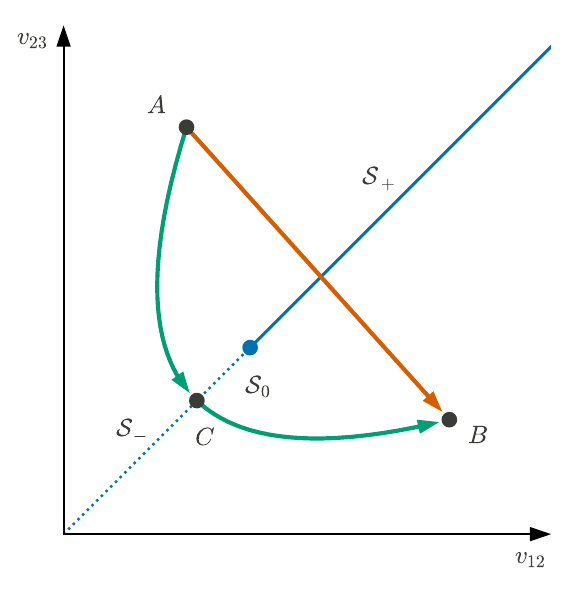}
        \caption{Two-dimensional slice of phase space.}
        \label{fig:NestedRootAnalyticStructureLocus}
    \end{subfigure}
    \hfill
    \begin{subfigure}[c]{0.48\textwidth}
        \centering
        \includegraphics[width=\textwidth]{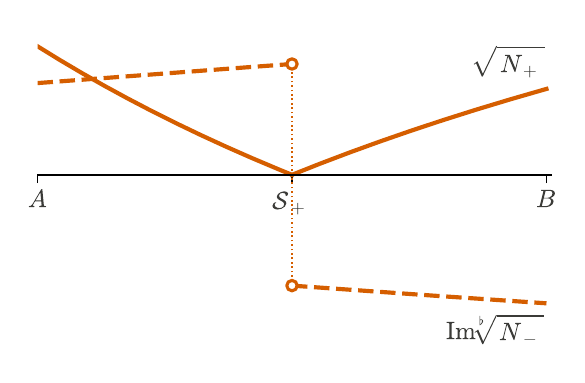}
        \caption{Nested roots along the straight line path between $A$ and $B$. Where the path crosses $\mathcal{S}_+$, $\sqrt{N_+}$ has a kink and $\sqrtalg{N_-}$ is discontinuous.}
        \label{fig:NestedRootAnalyticStructureKink}
    \end{subfigure}
    \caption{Structure of the non-analyticities of nested square roots on a two-dimensional slice of physical phase space, where all invariants other than $v_{12}$ and $v_{23}$ are fixed. The non-analyticities occur along $\mathcal{S}_+$, one half of the $v_{12}=v_{23}$ surface. This disqualifies integration paths that cross $\mathcal{S}_+$ (orange), while allowing those that bypass it (green).}
    \label{fig:NestedRootAnalyticStructure}
\end{figure}

Despite this comparatively simple behaviour of the inner and outer square roots
individually, it turns out that their interplay leads to a non-trivial analytic structure of the nested roots on physical phase space.
We sketch the general features in \cref{fig:NestedRootAnalyticStructure}. The
surface $v_{12} = v_{23}$ plays a special role in the analytic structure of the
nested roots, although this surface is not a Landau singularity and consequently
$v_{12} - v_{23}$ is not a letter in the alphabet.
Indeed, on this surface we have that
\begin{equation}
    \left.N_{\pm}\right|_{v_{12}=v_{23}}
    =
    \left.q^2\, (N_b\pm|N_b|)\right|_{v_{12}=v_{23}}.
\end{equation}
Thus, away from $\mathcal{S}_0$, exactly one of $N_+$ and $N_-$ vanishes on $v_{12}=v_{23}$. Which of the two vanishes is determined by the sign of $N_b$. It is therefore useful to distinguish the two regions
\begin{equation}
    \mathcal{S}_{\pm}
    =
    \{v_{12}=v_{23}\}\cap\{\pm N_b < 0\},
\end{equation}
where $N_{\pm}$ vanish respectively.
\Cref{fig:NestedRootAnalyticStructureLocus} shows these surfaces on a
two-dimensional slice of physical phase space, in which all invariants other
than $v_{12}$ and $v_{23}$ are held fixed. In this slice, the surface
$v_{12}=v_{23}$ becomes the diagonal. Since $\mathcal{S}_0$ has codimension two,
it reduces to a single point on the diagonal, which separates $\mathcal{S}_+$
(solid) from $\mathcal{S}_-$ (dotted).

While the vanishing of the nested roots on $\mathcal S_\pm$ is manifest, their
behaviour in a neighbourhood of these surfaces is more subtle.
We first analyse the behaviour of the two nested roots in a neighbourhood of $\mathcal{S}_+$.
Consider $\sqrt{N_+}$. By \cref{eq:NpNmProduct}, we can rewrite it in the form
\begin{equation}
    N_+
    =
    (v_{12}-v_{23})^2
    \frac{
        \mathcal{C}_1
        [q^2(v_{12} + v_{23} + 2 v_{13} + 4m_t^2)]^2
    }{N_-}.
\end{equation}
Given the sign properties of the various factors over physical phase space, this representation greatly simplifies taking the principal square root. Indeed, the only quantity that is not of definite sign is $v_{12} - v_{23}$ and we can write
\begin{equation}
    \sqrt{N_+}
    =
    |v_{12}-v_{23}|
        \frac{
            \sqrt{\mathcal{C}_1}
            [q^2(v_{12} + v_{23} + 2 v_{13} + 4m_t^2)]
            }{\sqrt{N_-}},
    \label{eq:RootNpKink}
\end{equation}
where we use the principal branch of the square root of $N_-$. This representation holds wherever $N_-\neq0$, in particular in a neighbourhood of $\mathcal{S}_+$.
As $\sqrt{N_-}$ is analytic along $\mathcal{S}_+$, the appearance of the absolute value in \cref{eq:RootNpKink} tells us that $\sqrt{N_+}$ is not smooth there. 
Thus, $\sqrt{N_+}$ develops a kink across $\mathcal{S}_+$.
Importantly, this behaviour flows into the behaviour of the algebraic root $\sqrtalg{N_-}$. Indeed,
combining its definition \cref{eq:RootNmDefinition} with \cref{eq:RootNpKink}, one finds
\begin{equation}
     \sqrtalg{N_-} = \operatorname{sign}(v_{12}-v_{23}) \sqrt{N_-}.
     \label{eq:algRootAsPrincipal}
\end{equation}
We therefore see that $\sqrtalg{N_-}$ is discontinuous across $\mathcal{S}_+$.

Importantly, the nested roots are much better behaved along $\mathcal{S}_{-}$.
As a point on $\mathcal{S}_-$ stays away from $\mathcal{S}_0$, both the
arguments of the inner and outer root when calculating $\sqrt{N_+}$ are away
from the branch locus. Hence, within a neighbourhood of a point on
$\mathcal{S}_-$, $\sqrt{N_+}$ is analytic. Thus, by \cref{eq:RootNmDefinition},
as $\sqrt{N_+}$ is non-zero, we see that $\sqrtalg{N_-}$ is also analytic along
$\mathcal{S}_{-}$.
To exemplify this behaviour, in \cref{fig:NestedRootAnalyticStructureKink}, we
depict the nested roots over the straight-line path from $A$ to $B$ (orange)
that crosses $\mathcal{S}_+$.
We note that along the path in \cref{fig:NestedRootAnalyticStructureLocus} that
passes through the point $C$ on $\mathcal{S}_-$, the square roots are perfectly
analytic.

Let us now consider how this analysis affects our solution of the pure
integrals.
In practice, paths that cross $\mathcal{S}_-$ are safe, while paths that cross
$\mathcal{S}_+$ must be treated with care.
This follows as the pure integrals, defined with the principal branch of $\sqrt{N_+}$, are themselves discontinuous across $\mathcal S_+$.
To see this, note that of the pure integrals, only $I_{1,64}$ and $I_{1,65}$ involve the nested roots. They are defined
as a combination of scalar Feynman integrals which are analytic on
$\mathcal{S}_+$ as it is not a Landau singularity. The combination itself, however, involves $\sqrt{N_+}$ and $\sqrtalg{N_-}$, which, by \cref{eq:RootNpKink,eq:algRootAsPrincipal}, causes one of the pure integrals to change sign across $\mathcal S_+$.
Along a path that crosses $\mathcal S_+$, \cref{eq:pentagonIT} instead yields a continuous function, which therefore no longer reproduces the pure integrals beyond the crossing.
While paths that cross $\mathcal{S}_+$ could in principle be treated
consistently,\footnote{A consistent treatment of such paths can be obtained by tracking the number of times $n$ that $\gamma$ has crossed $\mathcal{S}_+$, and redefining the square root $\sqrt{N_+}$ by a factor $(-1)^n$.
The result must carry this information: a phase-space point is no longer specified by the Mandelstam invariants alone, but also by the sign of $\sqrt{N_+}$, which flips across $\mathcal{S}_+$.
This ambiguity arises only in the pure integrals. It cancels in the rotation
matrix to the Feynman-integral basis, which, as expected, only depends on the
Mandelstam invariants.
}
we choose to avoid this class of complications by only considering paths $\gamma$
that do not cross $\mathcal{S}_+$.
In practice, the straight-line paths that we use in \cref{sec:numerics} start at
a point $\vec s_0\in\mathcal S_-$ and do not cross the surface $v_{12}=v_{23}$
again. In particular, they never cross $\mathcal{S}_+$.

%--------------------------------------------------------------------------------
\subsection[Weights 3 and 4: One-Fold Integral Representations]{Weights 3 and 4: One-Fold Integral Representations}
\label{sec:weight3}\label{sec:weights3And4}
%--------------------------------------------------------------------------------

At weights 3 and 4, we again choose a basis of pentagon functions modulo
products from components of the $\epsilon$ expansion of the Feynman integrals.
As is well known, in contrast to weights 1 and 2, the construction of representations of the special 
functions in terms of polylogarithms is a demanding task. We therefore follow the standard procedure of 
representing the functions as one-fold integrals along a path in phase space~\cite{Caron-Huot:2014lda}.
To this end, we introduce a path $\gamma(t)$ between an initial point $\vec{s}_0$ and a final point $\vec{s}_1$, reached at points $t = 0$ and $t=1$ respectively. That is,
\begin{equation}
\label{eq:path}
    \gamma(0) = \vec{s}_0, \qquad \gamma(1) = \vec{s}_1.
\end{equation}
The explicit construction of the path and the choice of the boundary point $\vec{s}_0$ are given in \cref{sec:numerics}.
For the moment, we will only require that
$\gamma(t)$ is a path within physical phase space that does not cross
$\mathcal{S}_+$ (see \cref{sec:nested_root}). The paths we use start at $\vec s_0\in\mathcal S_-$ and do not return to the surface $v_{12}=v_{23}$; this is used below at weight~4.
Along the path, we abbreviate
\begin{equation}
    W_j(t)\equiv W_j[\gamma(t)]\,,\qquad g_i^{(w)}(t)\equiv g_i^{(w)}(\gamma(t))\,,
    \label{eq:path_shorthand}
\end{equation}
so that, e.g., $g_i^{(w)}(0)=g_i^{(w)}(\vec s_0)$ and $W_j(1)=W_j[\vec s_1]$. Below, we first give the one-fold
integral representation at weight 3, and thereafter discuss the one-fold
integral representation at weight 4.

\paragraph{Weight 3:}
Given a path $\gamma(t)$, the linearly independent special functions at weight 3
can be written as a one-fold integral as
\begin{align}
    g_i^{(3)}(\vec{s}_1) = g_i^{(3)}(\vec{s}_0) + c_{ijk}^{(3)} \int_0^1 \mathrm{d}t \, \partial_t \log W_j(t)\; g_k^{(2)}(t),
    \label{eq:Weight3Abs}
\end{align}
where $c_{ijk}^{(3)}$ was defined in \cref{eq:pentagonDE}.
Naturally, the weight 3 pentagon functions $f_i^{(3)}$ are simply a subset of
the linearly independent special functions $g_j^{(3)}$ and so
\cref{eq:Weight3Abs} can be used to evaluate them.
Importantly, as $\gamma$ is a path entirely within the physical region, we can
use the $g_k^{(2)}$, expressed through the $f_j^{(2)}$, products of the
$f_j^{(1)}$ and $\zeta_2$ constructed in \cref{sec:weight1And2}, in the
integrand of \cref{eq:Weight3Abs}. Thus, only the boundary values
$g_i^{(3)}(\vec{s}_0)$ are required, which can be easily obtained from programs
such as \texttt{AMFlow}~\cite{Liu:2022chg,Huang:2026rjb}.

The representation in \cref{eq:Weight3Abs} is commonly used to compute pentagon
functions via numerical integration (see e.g.\
refs.~\cite{Caron-Huot:2014lda,Gehrmann:2018yef,Chicherin:2020oor,Chicherin:2021dyp,Abreu:2023rco}).
This is the approach that we will follow in \cref{sec:numerics}. As none of the
letters vanishes inside the physical region (\cref{sec:alphabet_phys}), very
little dedicated analytic work is required for the implementation of most weight
3 pentagon functions. However, at weight 3 we first encounter letters which
depend on the nested square root, which requires dedicated study.
In particular, issues arise when numerically evaluating a $\partial_t\log W_i(t)$ that involves the nested root.
Specifically, the analytic expression of such a derivative may involve poles at one of $v_{12} - v_{23}=0$, $\sqrt{N_+} = 0$ or $\sqrtalg{N_-}=0$.
Given that $v_{12}-v_{23}$ is not a letter, one expects such poles to be spurious.
In practice, however, finding a  representation of the $\mathrm{d}\log(W_i)$ in which all spurious poles are manifestly absent is non-trivial.
Nevertheless, it is relatively straightforward to construct multiple representations, each of which is manifestly regular on one of $\mathcal{S}_+$ or $\mathcal{S}_-$.

As an illustrative example, consider the letter $W_{115}$, which can be factorised into
\begin{equation}
    W_{115}^a
    =
    \Omega_1\!\left(q^2(v_{12}\!-\!v_{23}),N_+\right),
    \qquad
    W_{115}^b
    =
    \Omega_1\!\left(
    \frac{q^2(v_{12}\!-\!v_{23})(v_{45}\!+\!v_{13}\!+\!2m_t^2\!-\!q^2)}
         {v_{24}+v_{25}},
    N_+\right),
    \label{eq:W115abDef}
\end{equation}
where we define the manifestly odd combination through
\begin{equation}
    \Omega_1(w,R)
    =
    \frac{w-\sqrt{R}}{w+\sqrt{R}}\,.
    \label{eq:O1}
\end{equation}
Let us focus on the factor $W_{115}^a$. If we compute its $\mathrm{d}\log$ naively, we find
\begin{equation}
    \mathrm{d}\log(W_{115}^a)
    =
    \frac{
        2N_+\,\mathrm{d}\bigl[q^2(v_{12}-v_{23})\bigr]
        -q^2(v_{12}-v_{23})\,\mathrm{d}N_+
    }
    {
        \sqrt{N_+}
        \bigl[q^4(v_{12}-v_{23})^2-N_+\bigr]
    }\,.
\end{equation}
This representation is manifestly free of a pole on $\mathcal{S}_-$. On
$\mathcal{S}_+$, however, its denominator vanishes as
$\mathcal{O}((v_{12}-v_{23})^3)$, despite this singularity being spurious.
Alternatively, some ad-hoc manipulations allow us to rewrite the $\mathrm{d}\log$ form as
\begin{equation}
    \mathrm{d}\log(W_{115}^a)
    =
    \frac{
        2N_-\,\mathrm{d}\bigl[
        \sqrt{\mathcal{C}_1}(2q^2+v_{12}+v_{23}-2v_{45})
        \bigr]
        -
        \sqrt{\mathcal{C}_1}(2q^2+v_{12}+v_{23}-2v_{45})\,\mathrm{d}N_-
    }
    {
        \sqrtalg{N_-}
        \bigl[
        \mathcal{C}_1(2q^2+v_{12}+v_{23}-2v_{45})^2-N_-
        \bigr]
    }\,.
    \label{eq:W115aDlogSecond}
\end{equation}
In contrast to the previous representation, the form in
\cref{eq:W115aDlogSecond} is manifestly regular on $\mathcal{S}_+$.
Nevertheless, it contains a corresponding spurious pole on $\mathcal{S}_-$.
Thus, by choosing between the two representations according to the region of interest, 
the spurious singularities can be avoided.

In this example, we have therefore shown that the $\mathrm{d}\log$ form can be made regular throughout $v_{12}=v_{23}$, at the cost of using multiple equivalent representations. 
In practice, we perform an analogous ad-hoc analysis for each letter involving nested roots. Choosing the representation appropriate to the relevant branch locus removes spurious numerical singularities and substantially improves the stability of the numerical integration.

\paragraph{Weight 4:}
\label{sec:weight4}
At weight 4, we follow the same strategy outlined for weight 3 and we construct
an integral representation for the weight 4 functions.
Along the path $\gamma$, \cref{eq:pentagonIT} at weight 4 is an integral over the weight 3 functions.
In the notation of \cref{eq:path_shorthand}, we have:
\begin{align}
\label{eq:w4_transport}
g_i^{(4)}(\vec s_1) &= g_i^{(4)}(\vec s_0) + c^{(4)}_{ijk}\int_0^1 \mathrm{d}t\; \partial_t\log W_j(t)\;g_k^{(3)}(t)
\\\nonumber
&= g_i^{(4)}(\vec s_0) + c^{(4)}_{ijk}\int_0^1 \mathrm{d}t\; \partial_t\log W_j(t)
\left( g_k^{(3)}(\vec s_0) + c^{(3)}_{klm}\int_0^t \mathrm{d}s\; \partial_s\log W_l(s)\;g_m^{(2)}(s)\right).
\end{align}
This two-fold integral representation can be expressed in a one-fold
representation by exchanging the order of integration over $s$ and $t$ in the
second term:
\begin{equation}
    g_i^{(4)}(\vec s_1) = c^{(4)}_{ijk}c^{(3)}_{klm}\int_0^1 \mathrm{d}s \;\partial_s\log W_l(s)\;g_m^{(2)}(s)\,L_j(s)
    + c^{(4)}_{ijk}\, g_k^{(3)}(\vec s_0)\, L_j(0) + g_i^{(4)}(\vec s_0)\,,
    \label{eq:w4_impl_form}
\end{equation}
where we define the function $L_j$ as
\begin{equation}
L_j(s) = \int_s^1\mathrm{d}t\;\partial_t\log(W_j(t)).
\label{eq:LogarithmAntiderivative}
\end{equation}
As at weight 3, the weight 4 pentagon functions are elements $f_i^{(4)}$ of
$g^{(4)}$, so \cref{eq:w4_impl_form} gives them directly.

The integral in \cref{eq:LogarithmAntiderivative} can be written in closed form
in terms of logarithms of letters, provided their branches are chosen such that
they are continuous along every physical path.
There are therefore two subtleties to consider: choosing such branches, and
evaluating the letters themselves stably near $v_{12}=v_{23}$, which we will now
discuss.
In most cases, as described in the alphabet study in \cref{sec:alphabet_phys},
we can choose the branches in this way along any path over physical phase-space.
In particular, we can express $L_j$ as:
\begin{equation}
\label{eq:Lj}
    L_j(s)=
    \begin{cases}
    \log\left(\dfrac{W_j(1)}{W_j(s)}\right)\,, & W_j\in\mathcal A_\mathbb{R}\,,\\[2ex]
    \log\bigl(\sigma_j W_j(1)\bigr)-\log\bigl(\sigma_j W_j(s)\bigr)\,, & W_j\in\mathcal A_\mathbb{C}\setminus(\mathcal A_\mathbb{C}^\circ 
    %{}\cup\{W_{114}\}
    )\,,
    %\\[2ex] \log W_j(1)-\log W_j(s)\,, & W_j=W_{114}\,,
    \end{cases}
\end{equation}
with the principal branch of the logarithm. 
All three expressions are continuous
along every physical path, as follows from the behaviour of the letters over
phase space.
If $W_j\in \mathcal A_{\mathbb R}$, $W_j$ has a fixed sign over
phase space and the ratio is therefore positive throughout.
If $W_j\in \mathcal A_{\mathbb C}$, $W_j$ is a pure phase and the discussion
becomes more subtle. For $W_j\in\mathcal A_\mathbb{C}\setminus\mathcal
A_\mathbb{C}^\circ$, the image of the physical region misses one of the points
$W=\pm1$ (\cref{sec:alphabet_phys}). 
The sign $\sigma_j=\pm 1$ is then chosen for each one of these letters in order to avoid the crossing of the logarithmic branch-cut.
%For all of these letters but one, the missing point is $W=+1$, so that $-W_j$ never reaches the negative real axis.
%The exception is $W_{114}$, which equals $+1$ on $\mathcal S_-$ but never
%reaches $W=-1$, and for which we therefore use $\log W_{114}$ itself.
%
For the 6 letters $W_j\in\mathcal A_\mathbb{C}^\circ$ listed in
\cref{eq:circles}, the image covers the whole unit circle
(\cref{sec:alphabet_phys}), and \cref{eq:Lj} would have to be supplemented by
winding numbers.
Instead, for these letters we evaluate $L_j(s)$ directly as the integral of its
$\mathrm{d}\log$ form, \cref{eq:LogarithmAntiderivative}, using a Chebyshev
approximation~\cite{Abreu:2026vxw}, which we discuss further in \cref{sec:Chebyshev}.

A second subtlety again concerns the nested square roots. Unlike at weight 3, where only $\mathrm{d}\log$ forms enter, \cref{eq:Lj} requires the values of the letters themselves, which must be evaluated stably near $v_{12}=v_{23}$.
Let us return to our $W_{115}^a$ example. Its representation in \cref{eq:W115abDef} takes the indeterminate ``zero over zero'' form on $\mathcal{S}_+$ as both numerator and denominator vanish. This can be avoided by using the equivalent representation
\begin{equation}
    W_{115}^a
    =
    \Omega_1^\flat\!\left(
    (2q^2+v_{12}+v_{23}-2v_{45})\sqrt{\mathcal{C}_1},
    N_-
    \right),
    \label{eq:W115aSecondRep}
\end{equation}
where $\Omega_1^\flat$ denotes $\Omega_1$ making use of the algebraic square root of \cref{eq:RootNmDefinition}.
Importantly \cref{eq:W115aSecondRep} manifestly avoids the zero over zero behaviour on $\mathcal{S}_+$. Evaluating $W_{115}^a$ using this second representation therefore gives a substantially more stable result in the vicinity of $\mathcal{S}_+$. The same strategy can be applied to other letters as needed, for which we construct suitable alternative representations on an ad-hoc basis.

More generally, we find numerically that none of the $W_i$ vanish on $\mathcal{S}_\pm$. Thus, despite the apparent singular behaviour of some naive representations, the letters themselves are regular on $v_{12}=v_{23}$. We do not have an algebraic proof of this statement, but have verified it numerically for the letters relevant to our implementation. This property is central to the numerical stability of our approach.

%--------------------------------------------------------------------------------
\section[Numerical Evaluation]{Numerical Evaluation}
\label{sec:numerics}
%--------------------------------------------------------------------------------
In \cref{sec:basis} we discussed our analytic construction of a set of
pentagon functions to express the planar light-quark-loop integrals for $pp
\rightarrow t\overline{t}H$ at two loops. In this section we discuss our
implementation of their numerical evaluation. For weights one and two we simply
implement their analytic forms and evaluate them numerically. Most of our
discussion will therefore centre on weights three and four.

\subsection{Implementation of One-Fold Integrals}
\label{sec:Chebyshev}

The weight-three and weight-four functions are obtained from
\cref{eq:Weight3Abs,eq:w4_impl_form}.
To avoid the analyticity issues discussed in \cref{sec:weights3And4} we choose
a boundary point on the surface $\mathcal{S}_-$.
Concretely, we choose the rational phase-space point
\begin{equation}
\begin{split}
  \vec{s}_0 &=
  \{v_{12}, v_{23}, v_{34}, v_{45}, v_{15}, m_t^2, q^2\} \\
  &= \{\num{100000},\, \num{100000},\, \num{-150000}, \, \num{570000},\, \num{-170000}, \, (172)^2, \, (125)^2 \}\,,
  \label{eq:boundary-point-v2}
\end{split}
\end{equation}
in units of $\text{GeV}^2$, corresponding to $m_t=172$~GeV and $q=125$~GeV.
At this point we evaluate the canonical master integrals with
\amflow{}~\cite{Liu:2022chg,Huang:2026rjb} and extract the boundary values
$g_i^{(w)}(\vec s_0)$ to 70 digits.

We choose our integration path to be a straight line starting at $\vec{s}_0$ and
ending at $\vec{s}_1$, i.e.
\begin{equation}
  \gamma(t) = \vec s_0 + t\,\big(\vec s_1 - \vec s_0\big)\,,
  \qquad t\in[0,1]\,\text{.}
  \label{eq:straightPath-v2}
\end{equation}
Such a line never returns to the $v_{12} = v_{23}$ surface, and the nested roots
are analytic along it for $t>0$.
As is well-known, straight line paths such as \cref{eq:straightPath-v2} do not
always stay inside the physical region as they may violate the Gram constraint
$\Delta_5 < 0$.
For lines that do violate this constraint, we reach $\vec s_1$ in two
straight segments, $\vec s_0\to\vec s_I\to\vec s_1$, through an intermediate
physical point $\vec s_I$ generated using \rambo{}~\cite{Kleiss:1985gy}.
Importantly, we require $\vec{s}_I$ to lie on the same side of $v_{12} = v_{23}$
as $\vec{s}_1$, such that the two-segment path is guaranteed not to cross
$v_{12} = v_{23}$.
In practice, only $0.6\%$ of the points in our phase-space sample require a two-segment path.

To evaluate the one-fold integrals, we use the \texttt{PentagonFunctions++}
approach of tanh--sinh quadrature~\cite{Takahasi:1974}, implemented in
\texttt{Boost.Math}~\cite{BoostMath}.
Such a quadrature rule exhibits slow convergence when a singularity of the
integrand approaches the interior of the interval from the complex plane.
As no letter vanishes inside the physical region (\cref{sec:alphabet_phys}),
such singularities never lie on the path itself, but they can come close to it.
We took particular care in two cases.
The first is the inner root of the nested root.
Every letter that involves the nested roots depends on $\sqrt{N_b^2 - N_c}$ and since
\begin{equation}
  \mathrm{d}N_\pm = q^2\left(\mathrm{d}N_b \pm \frac{\mathrm{d}(N_b^2-N_c)}{2\sqrt{N_b^2-N_c}}\right)
  + \frac{\mathrm{d}q^2}{q^2}\,N_\pm\,,
  \label{eq:dNpm-v2}
\end{equation}
their $\mathrm{d}\log$ forms contain $\sqrt{N_b^2-N_c}$ in the denominator.
The second is the even letter $W_{34}$. Every even letter gives poles of its
$\mathrm{d}\log$ form at its zeros, but only $W_{34}$ produced zeros close to
the integration interval in our study.
Along a straight segment, both $N_b^2 - N_c$ and $W_{34}$ are polynomials of
degree six in $t$ with real coefficients, and we find that neither vanishes on
generic paths on physical phase space.
As already noted, both are of fixed sign over physical phase space and therefore
zeros near the segment come in complex-conjugate pairs $(t^\ast,\bar
t^{\,\ast})$.
This structure allows us to easily remedy the instabilities they induce.
For each segment, we compute the roots of both polynomials with the
Durand--Kerner method~\cite{Kerner:1966}. For every pair with
$\mathrm{Re}\,t^\ast\in(0,1)$ and $|\mathrm{Im}\,t^\ast|<\delta$, we split the
integration interval at $\mathrm{Re}\,t^\ast$. After the split, the nearly real
singularity lies close to an endpoint of each subinterval, where the tanh--sinh
rule handles it well. We use $\delta=0.1$.

A further issue where we place some care is in the numerical evaluation of the
nested roots themselves.
A first simple observation is that the evaluation of the nested root
$\sqrtalg{N_-}$ using the representation in \cref{eq:RootNmDefinition} can be
unstable in regions where $\sqrt{N_+}$ is small.
In practice this can be remedied by using \cref{eq:algRootAsPrincipal}, which
avoids the problematic division by a small $\sqrt{N_+}$.
Beyond this, a stability issue arises in the region $N_c \sim 0$, where a naive
evaluation of the nested roots can suffer from loss of precision.
Specifically, if $N_c$ is small, one has
\begin{equation}
    N_b \sim \sqrt{N_b^2-N_c}\,,
\end{equation}
and, depending on the sign of $N_b$, the corresponding expressions for $N_\pm$
can therefore involve a subtraction of nearly equal quantities.
For definiteness, consider the case $N_b>0$, since the case $N_b<0$ can be
obtained by symmetry. We can then replace the potentially unstable combination
by
\begin{equation}
    N_b-\sqrt{N_b^2-N_c}
    =
    \frac{N_c}{N_b+\sqrt{N_b^2-N_c}}\,.
\end{equation}
As $N_c$ is a simple, factorized polynomial expression and the denominator on
the right-hand side involves no cancellation, this representation is
substantially more stable than the direct evaluation of the expression on the
left.

Beyond these dedicated tricks, in order to further improve numerical stability
we find it useful to numerically evaluate some of the $\mathrm{d}\log$ forms in
higher precision.
Specifically, when the library is built with support for extended-precision
arithmetic, using the \textsc{QD} library~\cite{qdlibrary},
as for the quadruple-precision evaluations of \cref{sec:performance}, we
evaluate a subset of the $\mathrm{d}\log$ forms at the quadrature nodes in
quadruple or octuple precision, and convert the result back to the working
precision before it enters the integrand.
All other $\mathrm{d}\log$ forms, and the rest of the integrand, are evaluated
in the requested working precision.
Since only a few $\mathrm{d}\log$ forms are affected, the cost of this promotion
is small compared to the cost of promoting the entire pentagon function to
higher precision.

Finally, for the six letters of \cref{eq:circles} we evaluate $L_j$ directly
from its integral representation, \cref{eq:LogarithmAntiderivative}, as
motivated at the end of \cref{sec:weights3And4}.
For this we use the adaptive Chebyshev
approximation~\cite{Liu:2026cpf,Abreu:2026vxw,Liu:2026hdp}, following
ref.~\cite{Abreu:2026vxw}, which we have implemented in \texttt{C++} for all
three working precisions.
On $t\in[0,1]$ we approximate the integrand $\partial_t\log W_j(t)$ by a
piecewise Chebyshev interpolant, which can be integrated analytically. This
yields a piecewise polynomial approximation of $L_j(s)$ that can be evaluated at
any point of the path and so plays the role of a closed-form antiderivative.
Each piece starts with $9$ Chebyshev nodes, which are doubled up to $129$ until
the coefficients have decayed below the target accuracy; otherwise the piece is
bisected.
The interpolants are built once per phase-space point, before the outer
integration over $t$, so that at each quadrature node $L_j$ costs only a
polynomial evaluation.
As the $\mathrm{d}\log$ forms have no spurious singularities on physical phase
space, the construction converges without further intervention.

%--------------------------------------------------------------------------------
\subsection[Rescue System, Performance and Validation]{Rescue System, Performance and Validation}
\label{sec:performance}
%--------------------------------------------------------------------------------
In this section we discuss our precision rescue system, and the validation and
performance of our numerical implementation of the pentagon functions.

To validate the implementation and assess its performance, we study our implementation on a physical sample of phase-space points.
Our pool of $60000$ phase-space points is generated with
\textsc{Matrix}~\cite{Grazzini:2017mhc} for the subprocess $gg\to t\bar t H$ at
$\sqrt{s}=13.6$~TeV, distributed according to the LO phase-space weight.
Of
these, $10000$ have been generated for the inclusive cross section with standard
PDG values for the top quark and Higgs boson masses, $10000$ in the same setup but
probing the phase space in the tail of the distributions, and $40000$ by varying
the masses of the top quark and Higgs boson, one at a time, by $\pm 10$ GeV.
We validate our implementation by taking a random subset of $\mathcal{O}(100)$
points from this sample and compare the values of all pentagon functions to
those obtained by a high-precision numerical evaluation of the master integrals
using \amflow{}. We find excellent agreement between the two approaches.

Over such a physical phase space, a small set of evaluations lose a significant
amount of precision and we therefore implement a precision rescue system that
detects such phase space points and automatically repeats the calculation at
higher precision.
We detect unstable points by making use of two error measures on our results.
For definiteness, we discuss the case of weight 4 functions, and decompose the
expression in \cref{eq:w4_impl_form} into its three contributions, according to
\begin{equation}
 g_i^{(4)}(\vec s_1) = A_i + B_i +C_i\;,
\end{equation}
with
\begin{align}
    A_i &= c^{(4)}_{ijk}c^{(3)}_{klm}\int_0^1 \mathrm{d}s \;\partial_s\log W_l(s)\;g_m^{(2)}(s)\,L_j(s)\;, \\
B_i &= c^{(4)}_{ijk}\, g_k^{(3)}(\vec s_0)\, L_j(0)\;, \\
C_i&= g_i^{(4)}(\vec s_0)\,.
\end{align}
A similar expression holds for weight 3 functions, see \cref{eq:Weight3Abs}, where $B_i=0$.
One source of numerical error originates from the numerical integration to
obtain $A_i$, which is estimated by the integrator.
We denote the corresponding uncertainty estimation with $\Delta A_i$ and define
\begin{equation}
 \delta\sigma_1 = \frac{\Delta A_i}{|g_i^{(4)}|}\;.
\end{equation}
Another source of precision loss stems from large cancellations between the
three individual contributions in \cref{eq:w4_impl_form}.
To catch points in which large cancellations between the three contributions are
dominant we estimate their uncertainty by defining
\begin{equation}
 \delta \sigma_2 = \epsilon~\frac{|A_i| + |B_i| + |C_i|}{|A_i+B_i+C_i|}\;,
\end{equation}
where we have chosen $\epsilon \sim 10^{-8}$.
With these two error estimates in hand, we estimate the numerical error on a
given pentagon function as
\begin{equation}
 \Delta g_i^{(4)} = \max\{\delta \sigma_1,\delta \sigma_2\}\;.
\end{equation}
If the estimated error satisfies $\Delta g_i^{(4)} > 10^{-4}$, the rescue system is
triggered, which, for double and quad precision evaluations, will re-evaluate
the entire function in higher precision.

\begin{figure}[t]
    \centering
    \includegraphics[width=0.8\textwidth]{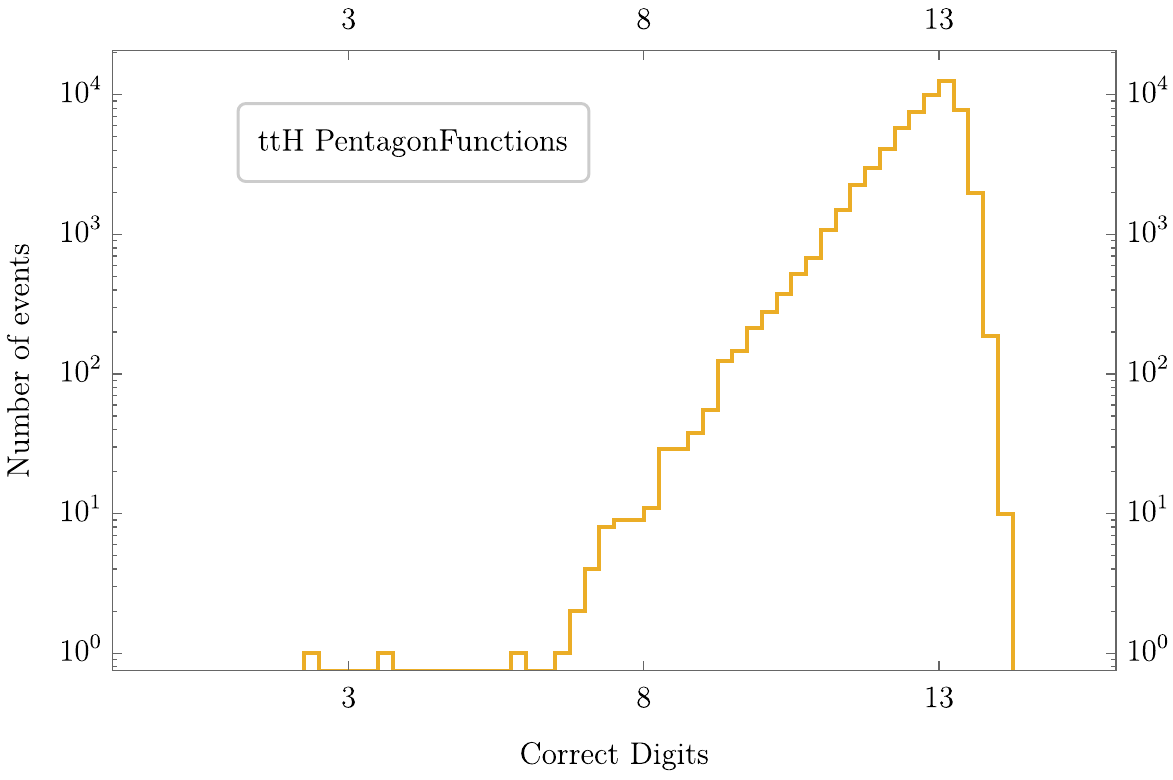}
    \caption[Distribution of the minimal number of correct digits]{Distribution of the minimal number of correct digits for all pentagon functions as measured by $E_r(\vec{s})$, given in \cref{eq:Er}.
    The special functions are evaluated on 60000 phase-space points.}
    \label{fig:precision}
\end{figure}

To address the performance and accuracy of our implementation, we
evaluate the entire sample of phase-space points and compare the results in
double precision with those obtained in quadruple precision.
We define
\begin{equation}
  E_r(\vec{s}) = \max_{n,\,i}\,
  \left|\,1 - \frac{f^{(n),\mathrm{quad}}_i(\vec{s})}
                   {f^{(n),\mathrm{double}}_i(\vec{s})}\right|\,,
  \label{eq:Er}
\end{equation}
where the superscripts ``quad'' and ``double'' denote evaluations in quadruple
and double precision, respectively.
Therefore, $E_r(\vec{s})$ is a measure for the numerical stability of our implementation. \Cref{fig:precision} shows the distribution of the number of correct
digits, $-\log_{10}E_r(\vec{s})$, over this sample. For the bulk of the
points we find at least $10$ correct digits. Only $1.1\%$ of the points have
fewer than $10$ correct digits leading us to conclude that our implementation is
suitable for phenomenological application.
The rescue system is triggered for $0.3\%$ of the points in
this sample. Its cost is included in the timings of \cref{tab:timing}.

Finally, to characterise the performance of the implementation, we report timings in
\cref{tab:timing}. These are measured on an \texttt{Intel Core i9-14900K}
with 32 threads over the same sample of phase-space points.

\begin{table}[t]
  \centering
  \begin{tabular}{lccc}
    \toprule
     & Median (ms) & Mean (ms) & Max (ms) \\
    \midrule
    double precision & 415 & 483 & 38827 \\
    quadruple precision & 35403 & 42554 & 2319508 \\
    \bottomrule
  \end{tabular}
  \caption{Timings for the evaluation of the full set of pentagon functions
    at a single phase-space point in double and quadruple precision. Timings are measured on an \texttt{Intel Core i9-14900K} with 32 threads.}
  \label{tab:timing}
\end{table}

\subsection{Ancillary Files}
\label{sec:ancillaryFiles}

In order to ease use of our analytic results and the pentagon functions code, we provide a series of ancillary files alongside this work. They are organised as follows:

\begin{itemize}
\item {\tt funcs.m}: a reference evaluation of the pentagon functions from \texttt{PentagonFunctions++} at the phase-space point given by $\mu^2 = 1$ and
    \begin{equation}
        s_4 = \{73781, 74098, -82315, 307009, -76978, 29929, 15625\},
    \end{equation} 
    in units of $\text{GeV}^2$.
\item {\tt letters.m}: provides the letters in terms of kinematical invariants including both the auxiliary letters $W_{i,X}$ and the four letters $W_{130}$, ..., $W_{133}$ that were not in $\mathrm{d}\log$ form in ref.~\cite{FebresCordero:2023pww}.    
\item {\tt pentagonBasis\_weighti.m}: provides the substitution which writes each master integral at weight $i$ in terms of pentagon functions.
\item {\tt roots.m}: expresses the square roots appearing in the alphabet in terms of Mandelstam invariants.
\item {\tt subAnalytic\_weighti.m}: provides the substitution rule which writes each element of the pentagon basis at weight $i$ in analytic form, as a function of the letters. 
\item {\tt benchmark.m}: notebook evaluating all the master integrals in the phase space point $s_4$, both using the output of the \texttt{PentagonFunctions++} library and the analytic expressions, comparing them.
\end{itemize}
%--------------------------------------------------------------------------------
\section{Summary and Outlook}
\label{sec:conclusions}
%--------------------------------------------------------------------------------

In this paper we have presented a basis of special functions, together with an
efficient numerical implementation, for the planar two-loop master integrals
entering the leading-colour amplitudes for $pp\to t\bar{t}H$ with a closed
light-quark loop. Starting from the $\epsilon$-factorised differential equations
of ref.~\cite{FebresCordero:2023pww}, we expressed the master
integrals in terms of a minimal set of pentagon functions whose symbols are
built from the alphabet of 152 letters.
Functions of weights one and two were obtained analytically in terms of
logarithms and dilogarithms while functions of weights three and four were
represented as one-fold integrals along paths in phase space.
The main new feature of this problem is that the canonical differential
equations involve nested square roots. We analysed their behaviour on the
physical region and showed that, on the hypersurface $v_{12}=v_{23}$, which is
not a Landau singularity, $\sqrt{N_+}$ develops a kink and $\sqrtalg{N_-}$ a
discontinuity.
To sidestep this difficulty, we place the boundary point of the
integration paths on this hypersurface and construct paths that do not return
to it.
We also found that natural representations of some letters, and of their $\mathrm{d}\log$
forms, show spurious zero-over-zero behaviour or spurious poles on this
hypersurface, even though the letters themselves are regular there.
We remove these by finding representations that manifest the lack of spurious
poles along our integration path.
In addition, six letters are pure phases
whose image over the physical region covers the whole unit circle, so that no
choice of branch cut makes the logarithm a global antiderivative of their
$\mathrm{d}\log$ forms.
Rather than tracking the number of times the integration path would cross the
branch cut of a logarithm or partitioning phase space, we construct these
antiderivatives numerically along each integration path with an adaptive
Chebyshev approximation~\cite{Abreu:2026vxw}.
We implemented the resulting evaluation in \texttt{C++}, as an extension of the
\texttt{PentagonFunctions++}
library\footnote{\href{https://gitlab.com/pentagon-functions/PentagonFunctions-cpp}{\texttt{https://gitlab.com/pentagon-functions/PentagonFunctions-cpp}}}~\cite{Chicherin:2020oor,Chicherin:2021dyp,Abreu:2023rco}.
We show that our approach is sufficiently stable over a sample of physical phase
space to be used in phenomenological applications. Moreover, we find that a
typical evaluation of the full set of functions takes around half a second.

As the difficulties we encountered stem from the algebraic structure of the
alphabet rather than from this particular process, we expect that the strategy,
namely paths adapted to the non-analytic loci of the roots, equivalent letter
representations chosen by region, and Chebyshev approximation for letters whose
phase covers the whole unit circle, will carry over to other multi-scale
problems with similar structure (see
e.g.~\cite{Badger:2024fgb,Becchetti:2025oyb,Gotzfried:2026def,Bera:2026qxx}).
Our results provide a building block for precise studies of $pp\to t\bar{t}H$ at
NNLO.
The same integrals also enter the two-loop amplitudes for $pp\to t\bar{t}Z$, so
our results are relevant beyond $t\bar{t}H$ production.
Since the evaluation is fast enough for phase-space integration, our results mark an
important stepping stone towards the full computation of the double virtual
contribution of the NNLO corrections to $pp \rightarrow t\overline{t}H$.

%--------------------------------------------------------------------------------
\section*{Acknowledgements}
%--------------------------------------------------------------------------------
%
We thank Stefan Kallweit for generating the phase-space sample used in our
studies. 
We thank Vasily Sotnikov for assistance with the \texttt{PentagonFunctions++} code.
We thank Sebastian P\"ogel for assistance in constructing $\mathrm{d}\log$ forms.
AI tools assisted in the final stages of editing and \texttt{C++}
implementation of this work.
MK acknowledges financial support through the DGAPA-PAPIIT grant IA103126 at UNAM, and the SECIHTI ``Ciencia Básica y de Frontera'' grant CBF-2025-I-615.
The work of SD received support from the FWO (contract No. 1227426N) and the European Union (ERC, MultiScaleAmp, Grant Agreement No. 101078449). Views and opinions expressed
are however those of the authors only and do not necessarily reflect those of the European
Union or the European Research Council Executive Agency. Neither the European Union
nor the granting authority can be held responsible for them.

%--------------------------------------------------------------------------------
\appendix
\section{The Alphabet}
\label{app:alphabet}
\subsection{Properties of Letters Over Phase Space}
\label{sec:letters_prop}

 In the following, we list the letters appearing in the alphabet according to
 their behaviour over phase-space. We report the properties of all letters,
 including both the relevant letters that we make use of in this work, and
 irrelevant letters that only arise at $\mathcal{O}(\epsilon^5)$.
\begin{itemize}
    \item letters with fixed sign:
    \begin{align}
        \mathcal A_\mathbb{R}=\{&W_1, W_2, W_3, W_4, W_5, W_6, W_8, W_9, W_{10}, W_{11}, W_{12}, W_{13}, W_{14}, W_{15}, W_{16}, W_{17},
        \\\nonumber&
        W_{18}, W_{19}, W_{20}, W_{21}, W_{22}, W_{23}, W_{24}, W_{25}, W_{26}, W_{27}, W_{28}, W_{29}
        , W_{30}, W_{31}, W_{32},
        \\\nonumber&
        W_{33}, W_{34}, W_{35}, W_{36}, W_{37}, W_{38}, W_{39}, W_{40}, W_{41}, W_{42}, W_{43}, W_{44}, W_{45}, W_{46}, W_{47},
        \\\nonumber&
        W_{48}, W_{49}, W_{50}, W_{51}, W_{52}, W_{53}, W_{54}   , W_{58}, W_{61}, W_{62}, W_{63}, W_{64}, W_{65}, W_{66}, W_{67},
        \\\nonumber&
        W_{68}, W_{69}, W_{76}, W_{77}, W_{78}, W_{79}, W_{80}, W_{81}, W_{82}, W_{83}, W_{84}, W_{85}, W_{86}, W_{87}, W_{88},\\\nonumber&
        W_{89}, W_{95}, W_{97}, W_{98}, W_{104}, W_{105}, W_{109}, W_{110}, W_{113}, W_{115}, W_{117}, W_{119}, W_{120},
        \\\nonumber&
        W_{122}, W_{123}, W_{126}, W_{129}, W_{130}, W_{131}, W_{134}^2, W_{135}^2, W_{136}^2, W_{137}^2, W_{138}^2, W_{139}^2,
        \\\nonumber&
        W_{140}^2, W_{141}^2, W_{142}^2, W_{143}^2, W_{144}, W_{145}, W_{146}, W_{149}, W_{152}\}\,.
    \end{align}
    We note that in ref.~\cite{FebresCordero:2023pww} the letters $W_{134},\ldots,W_{143}$ are the square roots $\sqrt{\mathcal C_1}$, $\sqrt{\Delta_3^{(1)}},\allowbreak\ldots,\allowbreak\sqrt{\Delta_3^{(5)}}$, $\sqrt{\mathcal C_2}$, $\sqrt{\Delta_5}$, $\sqrt{r_2}$ and $\sqrt{r_3}$. We use their squares, which are polynomials of fixed sign.
    \item letters that are a pure phase:
    \begin{align}
        \mathcal A_\mathbb{C}=\{&W_7, W_{55}, W_{56}, W_{57}, W_{59}, \underline{W_{60}}, W_{70}, W_{71}, W_{72}, W_{73}, W_{74}, \underline{W_{75}},
        \\\nonumber&
        W_{90}, W_{91}, W_{92}, W_{93}, W_{94}, W_{96}, W_{99}, W_{100}, W_{101}, W_{102}, W_{103}, W_{106}, W_{107},
        \\\nonumber&
        W_{108}, W_{111}, W_{112}, W_{114}, W_{116}, \underline{W_{118}}, W_{121}, W_{124}, W_{125}, W_{127}, \underline{W_{128}},
        \\\nonumber&
        \underline{W_{132}}, \underline{W_{133}}, W_{147}, W_{148}, W_{150}, W_{151}\}\;.
    \end{align}
    The underlined letters are elements of $\mathcal A_\mathbb{C}^\circ$, the ones whose image over the physical region covers the full unit circle in the complex plane, making it impossible to define $\log(\underline{W_i})$ in such a way that a physical path cannot cross its branch-cut, as described in \cref{sec:alphabet_phys}.
\end{itemize}

\subsection{Auxiliary Letters}
\label{sec:aux_letters}

To construct analytic expressions for the weight-2 functions we introduced an
additional set of auxiliary letters, which further breaks down the original
alphabet.
The full set of auxiliary letters that we employ is defined through
\begin{align}
    W_7=&\frac{W_{7,A}}{W_{7,B}}\,;&W_{37}=&\frac{W_{37,A}}{W_{37,B}}\,;&W_{38}=&\frac{W_{38,A}}{W_{38,B}}\,;
    \\\nonumber
    W_{45}=&\frac{W_{45,A}}{W_{45,B}}\,;&W_{46}=&\frac{W_{46,A}}{W_{46,B}}\,;&W_{55}=&\frac{W_{55,A}}{W_{55,B}}\,;
    \\\nonumber
    W_{99}=&\frac{W_{99,A} W_{99,B}}{W_{99,C}W_{99,D}}\,;& W_{100}=&\frac{W_{100,A} W_{100,B}}{W_{100,C}W_{100,D}}\,; &W_{101}=&\frac{W_{101,A} W_{101,B}}{W_{101,C}W_{101,D}}\,;
\end{align}
where
\begin{align}
 W_{7,A}= & q^2+\sqrtC\,; &  W_{7,B}= & q^2-\sqrtC\,;
 \\\nonumber
 W_{37,A}= & 2\mTsq+v_{12}-\sqrtG\,; &  W_{37,B}= &  2\mTsq+v_{12}+\sqrtG\,;
 \\\nonumber
 W_{38,A}= & 2\mTsq+v_{23}-\sqrtGG\,; &  W_{38,B}= &  2\mTsq+v_{23}+\sqrtGG\,;
 \\\nonumber
 W_{45,A}= & q^2+v_{15}-v_{34}+\sqrtGfv\,; &  W_{45,B}= &   q^2+v_{15}-v_{34}-\sqrtGfv\,;
 \\\nonumber
 W_{46,A}= & q^2-v_{15}+v_{34}+\sqrtGfv\,; &  W_{46,B}= &   q^2-v_{15}+v_{34}-\sqrtGfv \,;
  \\\nonumber
 W_{55,A}= &2\mTsq v_{34}+v_{15}(q^2-2\mTsq-\sqrtC)\,; &  W_{55,B}= & 2\mTsq v_{34}+v_{15}(q^2-2\mTsq+\sqrtC) \,;
 \\\nonumber
 W_{99,A}= & -q^2-v_{12}-\sqrtC-\sqrtG\,; &  W_{99,B}= &   -q^2-v_{12}+\sqrtC+\sqrtG \,;
  \\\nonumber
 W_{99,C}= & -q^2-v_{12}+\sqrtC-\sqrtG\,; &  W_{99,D}= &   -q^2-v_{12}-\sqrtC+\sqrtG \,;
\\\nonumber
 W_{100,A}= & -q^2-v_{23}-\sqrtC-\sqrtGG\,; &  W_{100,B}= &   -q^2-v_{23}+\sqrtC+\sqrtGG \,;
  \\\nonumber
 W_{100,C}= & -q^2-v_{23}+\sqrtC-\sqrtGG\,; &  W_{100,D}= &   -q^2-v_{23}-\sqrtC+\sqrtGG \,;
 \\\nonumber
 W_{101,A}= & v_{34}-v_{15}-\sqrtC-\sqrtGfv\,; &  W_{101,B}= & v_{34}-v_{15}+\sqrtC+\sqrtGfv\,;
  \\\nonumber
 W_{101,C}= & v_{34}-v_{15}+\sqrtC-\sqrtGfv\,; &  W_{101,D}= & v_{34}-v_{15}-\sqrtC+\sqrtGfv\,.
\end{align}
We note that, similar to \cref{{eq:auxiliaryDlogRelationExample}},
$\mathrm{d}\log$s of auxiliary letters can be expressed in terms of
$\mathrm{d}\log$s of alphabet elements.

%--------------------------------------------------------------------------------
\bibliographystyle{JHEP}
\bibliography{ttHfuncs}

@article{ATLAS:2025eii,
    author = "Aad, Georges and others",
    collaboration = "ATLAS, CMS",
    title = "{Highlights of the HL-LHC physics projections by ATLAS and CMS}",
    eprint = "2504.00672",
    archivePrefix = "arXiv",
    primaryClass = "hep-ex",
    reportNumber = "ATL-PHYS-PUB-2025-018, CMS-HIG-25-002",
    month = "4",
    year = "2025"
}

@article{CMS:2018uxb,
    author = "Sirunyan, Albert M and others",
    collaboration = "CMS",
    title = "{Observation of $\mathrm{t\overline{t}}$H production}",
    eprint = "1804.02610",
    archivePrefix = "arXiv",
    primaryClass = "hep-ex",
    reportNumber = "CMS-HIG-17-035, CERN-EP-2018-064",
    doi = "10.1103/PhysRevLett.120.231801",
    journal = "Phys. Rev. Lett.",
    volume = "120",
    number = "23",
    pages = "231801",
    year = "2018"
}

@article{ATLAS:2018mme,
    author = "Aaboud, M. and others",
    collaboration = "ATLAS",
    title = "{Observation of Higgs boson production in association with a top quark pair at the LHC with the ATLAS detector}",
    eprint = "1806.00425",
    archivePrefix = "arXiv",
    primaryClass = "hep-ex",
    reportNumber = "CERN-EP-2018-138",
    doi = "10.1016/j.physletb.2018.07.035",
    journal = "Phys. Lett. B",
    volume = "784",
    pages = "173--191",
    year = "2018"
}

@article{Catani:2021cbl,
    author = "Catani, Stefano and Fabre, Ignacio and Grazzini, Massimiliano and Kallweit, Stefan",
    title = "{${t {{\bar{t}}}H}$ production at NNLO: the flavour off-diagonal channels}",
    eprint = "2102.03256",
    archivePrefix = "arXiv",
    primaryClass = "hep-ph",
    reportNumber = "ZU-TH 3/21, ICAS 061/21",
    doi = "10.1140/epjc/s10052-021-09247-w",
    journal = "Eur. Phys. J. C",
    volume = "81",
    number = "6",
    pages = "491",
    year = "2021"
}

@article{Catani:2022mfv,
    author = "Catani, Stefano and Devoto, Simone and Grazzini, Massimiliano and Kallweit, Stefan and Mazzitelli, Javier and Savoini, Chiara",
    title = "{Higgs Boson Production in Association with a Top-Antitop Quark Pair in Next-to-Next-to-Leading Order QCD}",
    eprint = "2210.07846",
    archivePrefix = "arXiv",
    primaryClass = "hep-ph",
    reportNumber = "TIF-UNIMI-2022-15, ZU-TH 46/22, MPP-2022-128, PSI-PR-22-30",
    doi = "10.1103/PhysRevLett.130.111902",
    journal = "Phys. Rev. Lett.",
    volume = "130",
    number = "11",
    pages = "111902",
    year = "2023"
}

@article{Devoto:2024nhl,
    author = "Devoto, Simone and Grazzini, Massimiliano and Kallweit, Stefan and Mazzitelli, Javier and Savoini, Chiara",
    title = "{Precise predictions for $ t\overline{t}H $ production at the LHC: inclusive cross section and differential distributions}",
    eprint = "2411.15340",
    archivePrefix = "arXiv",
    primaryClass = "hep-ph",
    reportNumber = "ZU-TH-57/24, PSI-PR-24-22, TUM-HEP-1538/24",
    doi = "10.1007/JHEP03(2025)189",
    journal = "JHEP",
    volume = "03",
    pages = "189",
    year = "2025"
}

@article{Balsach:2025jcw,
    author = "Balsach, Roger and others",
    title = "{State-of-the-art cross sections for $t\bar t H$: NNLO predictions matched with NNLL resummation and EW corrections}",
    eprint = "2503.15043",
    archivePrefix = "arXiv",
    primaryClass = "hep-ph",
    reportNumber = "LHCHWG-2025-001, IPPP/25/01, TUM-HEP-1549/25, UWThPh 2024-25, TTK-25-01, P3H-25-001, TIF-UNIMI-2025-3, MS-TP-25-02, PSI-PR-25-03, ZU-TH 08/25",
    doi = "10.21468/SciPostPhysCommRep.10",
    month = "3",
    year = "2025"
}

@article{Biello:2026nhj,
    author = "Biello, Christian and Savoini, Chiara and Signorile-Signorile, Chiara and Wiesemann, Marius",
    title = "{Next-to-next-to-leading order event generation for $ t\overline{t}H $ production with approximate two-loop amplitude}",
    eprint = "2603.06143",
    archivePrefix = "arXiv",
    primaryClass = "hep-ph",
    reportNumber = "CERN-TH-2026-031, MPP-2026-22, TUM-HEP-1595/26",
    doi = "10.1007/JHEP06(2026)070",
    journal = "JHEP",
    volume = "06",
    pages = "070",
    year = "2026"
}

@article{FebresCordero:2023pww,
    author = "Febres Cordero, F. and Figueiredo, G. and Kraus, M. and Page, B. and Reina, L.",
    title = "{Two-loop master integrals for leading-color $ pp\to t\overline{t}H $ amplitudes with a light-quark loop}",
    eprint = "2312.08131",
    archivePrefix = "arXiv",
    primaryClass = "hep-ph",
    reportNumber = "CERN-TH-2023-240",
    doi = "10.1007/JHEP07(2024)084",
    journal = "JHEP",
    volume = "07",
    pages = "084",
    year = "2024"
}

@article{Duhr:2011zq,
    author = "Duhr, Claude and Gangl, Herbert and Rhodes, John R.",
    title = "{From polygons and symbols to polylogarithmic functions}",
    eprint = "1110.0458",
    archivePrefix = "arXiv",
    primaryClass = "math-ph",
    reportNumber = "IPPP-11-56, DCPT-11-112",
    doi = "10.1007/JHEP10(2012)075",
    journal = "JHEP",
    volume = "10",
    pages = "075",
    year = "2012"
}

@article{Chicherin:2021dyp,
    author = "Chicherin, Dmitry and Sotnikov, Vasily and Zoia, Simone",
    title = "{Pentagon functions for one-mass planar scattering amplitudes}",
    eprint = "2110.10111",
    archivePrefix = "arXiv",
    primaryClass = "hep-ph",
    reportNumber = "LAPTH-041/21, MPP-2021-182",
    doi = "10.1007/JHEP01(2022)096",
    journal = "JHEP",
    volume = "01",
    pages = "096",
    year = "2022"
}

@article{Abreu:2026vxw,
    author = "Abreu, Samuel and Guerreiro, Afonso and Page, Ben",
    title = "{Chebyshev Approximations of Feynman Integrals for Collider Physics}",
    eprint = "2607.02411",
    archivePrefix = "arXiv",
    primaryClass = "hep-ph",
    month = "7",
    year = "2026"
}

@article{Liu:2022chg,
    author = "Liu, Xiao and Ma, Yan-Qing",
    title = "{AMFlow: A Mathematica package for Feynman integrals computation via auxiliary mass flow}",
    eprint = "2201.11669",
    archivePrefix = "arXiv",
    primaryClass = "hep-ph",
    doi = "10.1016/j.cpc.2022.108565",
    journal = "Comput. Phys. Commun.",
    volume = "283",
    pages = "108565",
    year = "2023"
}

@article{Huang:2026rjb,
    author = "Huang, Rui-Jun and Liu, Xiao and Ma, Yan-Qing",
    title = "{AMFlow 2.0: significant algorithmic and software improvements for Feynman integral evaluation}",
    eprint = "2607.08477",
    archivePrefix = "arXiv",
    primaryClass = "hep-ph",
    month = "7",
    year = "2026"
}

@article{Liu:2026cpf,
    author = "Liu, Yuanche and Zhang, Yang",
    title = "{CHESS: CHEbyshev pSeudo-Spectral transport for Feynman integral differential equations}",
    eprint = "2606.26691",
    archivePrefix = "arXiv",
    primaryClass = "hep-ph",
    reportNumber = "USTC-ICTS/PCFT-26-42",
    month = "6",
    year = "2026"
}

@article{Liu:2026hdp,
    author = "Liu, Yuanche and Matija{\v{s}}i{\'c}, Antonela and Peraro, Tiziano and Xu, Yingxuan and Yang, Zihua and Zhang, Yang",
    title = "{Two-loop six-point planar massless feynman integrals to higher {\ensuremath{\epsilon}} orders}",
    eprint = "2603.16831",
    archivePrefix = "arXiv",
    primaryClass = "hep-ph",
    reportNumber = "TTP26-009, P3H-26-020, USTC-ICTS/PCFT-26-19",
    doi = "10.1007/JHEP07(2026)223",
    journal = "JHEP",
    volume = "07",
    pages = "223",
    year = "2026"
}

@unpublished{PagePoegelTrager,
  author = {Ben Page and Sebastian Poegel},
  title  = {\texttt{Trager}: a package for algebraic integration for Feynman integrals},
  note   = {To appear},
}

@article{Agarwal:2024jyq,
    author = "Agarwal, Bakul and Heinrich, Gudrun and Jones, Stephen P. and Kerner, Matthias and Klein, Sven Yannick and Lang, Jannis and Magerya, Vitaly and Olsson, Anton",
    title = "{Two-loop amplitudes for $ t\overline{t}H $ production: the quark-initiated N$_{f}$-part}",
    eprint = "2402.03301",
    archivePrefix = "arXiv",
    primaryClass = "hep-ph",
    reportNumber = "IPPP/24/03, KA-TP-02-2024, P3H-24-007, TTK-24-03",
    doi = "10.1007/JHEP05(2024)013",
    journal = "JHEP",
    volume = "05",
    pages = "013",
    year = "2024",
    note = "[Erratum: JHEP 06, 142 (2024)]"
}

@article{Wang:2024pmv,
    author = "Wang, Guoxing and Xia, Tianya and Yang, Li Lin and Ye, Xiaoping",
    title = "{Two-loop QCD amplitudes for $ t\overline{t}H $ production from boosted limit}",
    eprint = "2402.00431",
    archivePrefix = "arXiv",
    primaryClass = "hep-ph",
    doi = "10.1007/JHEP07(2024)121",
    journal = "JHEP",
    volume = "07",
    pages = "121",
    year = "2024"
}

@article{Hidding:2020ytt,
    author = "Hidding, Martijn",
    title = "{DiffExp, a Mathematica package for computing Feynman integrals in terms of one-dimensional series expansions}",
    eprint = "2006.05510",
    archivePrefix = "arXiv",
    primaryClass = "hep-ph",
    doi = "10.1016/j.cpc.2021.108125",
    journal = "Comput. Phys. Commun.",
    volume = "269",
    pages = "108125",
    year = "2021"
}

@inproceedings{qdlibrary,
  author    = {Hida, Yozo and Li, Xiaoye S. and Bailey, David H.},
  title     = {Algorithms for Quad-Double Precision Floating Point Arithmetic},
  booktitle = {Proceedings 15th IEEE Symposium on Computer Arithmetic. ARITH-15 2001},
  address   = {Vail, CO, USA},
  year      = {2001},
  pages     = {155--162},
  doi       = {10.1109/ARITH.2001.930115}
}

@article{Kleiss:1985gy,
    author = "Kleiss, R. and Stirling, W. James and Ellis, S. D.",
    title = "{A New Monte Carlo Treatment of Multiparticle Phase Space at High-energies}",
    reportNumber = "CERN-TH-4299/85",
    doi = "10.1016/0010-4655(86)90119-0",
    journal = "Comput. Phys. Commun.",
    volume = "40",
    pages = "359",
    year = "1986"
}

@article{Gotzfried:2026def,
    author = {G{\"o}tzfried, Linus and von Manteuffel, Andreas},
    title = "{Two Feynman integral families for vector boson fusion at NNLO QCD}",
    eprint = "2609.37990",
    archivePrefix = "arXiv",
    primaryClass = "hep-ph",
    month = "9",
    year = "2026"
}

@article{Badger:2024fgb,
    author = "Badger, Simon and Becchetti, Matteo and Giraudo, Nicol{\`o} and Zoia, Simone",
    title = "{Two-loop integrals for $ t\overline{t} $+jet production at hadron colliders in the leading colour approximation}",
    eprint = "2404.12325",
    archivePrefix = "arXiv",
    primaryClass = "hep-ph",
    reportNumber = "CERN-TH-2024-048, ZU-TH 22/24",
    doi = "10.1007/JHEP07(2024)073",
    journal = "JHEP",
    volume = "07",
    pages = "073",
    year = "2024"
}

@article{Becchetti:2025oyb,
    author = "Becchetti, Matteo and Dlapa, Christoph and Zoia, Simone",
    title = "{Canonical differential equations for the elliptic two-loop five-point integral family relevant to tt{\textasciimacron}+jet production at leading color}",
    eprint = "2503.03603",
    archivePrefix = "arXiv",
    primaryClass = "hep-th",
    reportNumber = "DESY 25-029, ZU-TH 13/25",
    doi = "10.1103/zt4w-c1jk",
    journal = "Phys. Rev. D",
    volume = "112",
    number = "3",
    pages = "L031501",
    year = "2025"
}

@article{Becchetti:2025qlu,
    author = "Becchetti, Matteo and Canko, Dhimiter and Chestnov, Vsevolod and Peraro, Tiziano and Pozzoli, Mattia and Zoia, Simone",
    title = "{Two-loop Feynman integrals for leading colour $ t\overline{t}W $ production at hadron colliders}",
    eprint = "2504.13011",
    archivePrefix = "arXiv",
    primaryClass = "hep-ph",
    reportNumber = "ZU-TH 28/25",
    doi = "10.1007/JHEP07(2025)001",
    journal = "JHEP",
    volume = "07",
    pages = "001",
    year = "2025"
}

@article{Chicherin:2020oor,
    author = "Chicherin, Dmitry and Sotnikov, Vasily",
    title = "{Pentagon Functions for Scattering of Five Massless Particles}",
    eprint = "2009.07803",
    archivePrefix = "arXiv",
    primaryClass = "hep-ph",
    reportNumber = "MPP-2020-171",
    doi = "10.1007/JHEP12(2020)167",
    journal = "JHEP",
    volume = "20",
    pages = "167",
    year = "2020"
}

@article{Abreu:2023rco,
    author = "Abreu, Samuel and Chicherin, Dmitry and Ita, Harald and Page, Ben and Sotnikov, Vasily and Tschernow, Wladimir and Zoia, Simone",
    title = "{All Two-Loop Feynman Integrals for Five-Point One-Mass Scattering}",
    eprint = "2306.15431",
    archivePrefix = "arXiv",
    primaryClass = "hep-ph",
    reportNumber = "CERN-TH-2023-119",
    doi = "10.1103/PhysRevLett.132.141601",
    journal = "Phys. Rev. Lett.",
    volume = "132",
    number = "14",
    pages = "141601",
    year = "2024"
}

@article{Gehrmann:2018yef,
    author = "Gehrmann, T. and Henn, J. M. and Lo Presti, N. A.",
    title = "{Pentagon functions for massless planar scattering amplitudes}",
    eprint = "1807.09812",
    archivePrefix = "arXiv",
    primaryClass = "hep-ph",
    reportNumber = "ZU-TH-25-18, ZU-TH 25/18, MPP-2018-180, MITP-18-067, IPPP-18-59, MITP/18-067, IPPP/18/59",
    doi = "10.1007/JHEP10(2018)103",
    journal = "JHEP",
    volume = "10",
    pages = "103",
    year = "2018"
}

@article{Badger:2025ljy,
    author = "Badger, Simon and Becchetti, Matteo and Brancaccio, Colomba and Czakon, Micha{\l} and Hartanto, Heribertus Bayu and Poncelet, Rene and Zoia, Simone",
    title = "{Double virtual QCD corrections to $ t\overline{t} $+jet production at the LHC}",
    eprint = "2511.11424",
    archivePrefix = "arXiv",
    primaryClass = "hep-ph",
    reportNumber = "P3H-25-092, TTK-25-38, ZU-TH 74/25, IFJPAN-IV-2025-23",
    doi = "10.1007/JHEP05(2026)044",
    journal = "JHEP",
    volume = "05",
    pages = "044",
    year = "2026"
}

@article{PetitRosas:2025xhm,
    author = "Petit Ros{\`a}s, Pau and Torres Bobadilla, William J.",
    title = "{Fast evaluation of Feynman integrals for Monte Carlo generators}",
    eprint = "2507.12548",
    archivePrefix = "arXiv",
    primaryClass = "hep-ph",
    doi = "10.1007/JHEP09(2025)210",
    journal = "JHEP",
    volume = "09",
    pages = "210",
    year = "2025"
}

@article{Czakon:2026tog,
    author = "Czakon, Micha{\l} and Tancredi, Lorenzo",
    title = "{Solution of Canonical Differential Equations for Integrals on Arbitrary Geometries}",
    eprint = "2606.30354",
    archivePrefix = "arXiv",
    primaryClass = "hep-ph",
    reportNumber = "P3H-26-047,TTK-26-18,TUM-HEP-1607/26",
    month = "6",
    year = "2026"
}

@article{Chen:2026pyw,
    author = "Chen, Xiang and Zhang, Hantian",
    title = "{Feynman Integrals Meet Second-Order Partial Differential Equations}",
    eprint = "2607.18406",
    archivePrefix = "arXiv",
    primaryClass = "hep-ph",
    reportNumber = "CERN-TH-2026-170, ZU-TH 25/26",
    month = "7",
    year = "2026"
}

@article{Moriello:2019yhu,
    author = "Moriello, Francesco",
    title = "{Generalised power series expansions for the elliptic planar families of Higgs + jet production at two loops}",
    eprint = "1907.13234",
    archivePrefix = "arXiv",
    primaryClass = "hep-ph",
    doi = "10.1007/JHEP01(2020)150",
    journal = "JHEP",
    volume = "01",
    pages = "150",
    year = "2020"
}

@article{Caron-Huot:2014lda,
    author = "Caron-Huot, Simon and Henn, Johannes M.",
    title = "{Iterative structure of finite loop integrals}",
    eprint = "1404.2922",
    archivePrefix = "arXiv",
    primaryClass = "hep-th",
    doi = "10.1007/JHEP06(2014)114",
    journal = "JHEP",
    volume = "06",
    pages = "114",
    year = "2014"
}

@article{Armadillo:2022ugh,
    author = "Armadillo, Tommaso and Bonciani, Roberto and Devoto, Simone and Rana, Narayan and Vicini, Alessandro",
    title = "{Evaluation of Feynman integrals with arbitrary complex masses via series expansions}",
    eprint = "2205.03345",
    archivePrefix = "arXiv",
    primaryClass = "hep-ph",
    doi = "10.1016/j.cpc.2022.108545",
    journal = "Comput. Phys. Commun.",
    volume = "282",
    pages = "108545",
    year = "2023"
}

@article{Lee:2017qql,
    author = "Lee, Roman N. and Smirnov, Alexander V. and Smirnov, Vladimir A.",
    title = "{Solving differential equations for Feynman integrals by expansions near singular points}",
    eprint = "1709.07525",
    archivePrefix = "arXiv",
    primaryClass = "hep-ph",
    reportNumber = "TTP16-055, MITP-16-144",
    doi = "10.1007/JHEP03(2018)008",
    journal = "JHEP",
    volume = "03",
    pages = "008",
    year = "2018"
}

@article{Liu:2018dmc,
    author = "Liu, Xiao and Ma, Yan-Qing",
    title = "{Determining arbitrary Feynman integrals by vacuum integrals}",
    eprint = "1801.10523",
    archivePrefix = "arXiv",
    primaryClass = "hep-ph",
    doi = "10.1103/PhysRevD.99.071501",
    journal = "Phys. Rev. D",
    volume = "99",
    number = "7",
    pages = "071501",
    year = "2019"
}

@article{Liu:2017jxz,
    author = "Liu, Xiao and Ma, Yan-Qing and Wang, Chen-Yu",
    title = "{A Systematic and Efficient Method to Compute Multi-loop Master Integrals}",
    eprint = "1711.09572",
    archivePrefix = "arXiv",
    primaryClass = "hep-ph",
    doi = "10.1016/j.physletb.2018.02.026",
    journal = "Phys. Lett. B",
    volume = "779",
    pages = "353--357",
    year = "2018"
}

@article{Gehrmann:2024tds,
    author = "Gehrmann, Thomas and Henn, Johannes and Jakub{\v{c}}{\'\i}k, Petr and Lim, Jungwon and Mella, Cesare Carlo and Syrrakos, Nikolaos and Tancredi, Lorenzo and Torres Bobadilla, William J.",
    title = "{Graded transcendental functions: an application to four-point amplitudes with one off-shell leg}",
    eprint = "2410.19088",
    archivePrefix = "arXiv",
    primaryClass = "hep-th",
    reportNumber = "TUM-HEP 1529/24, MPP-2024-193, ZU-TH 52/24",
    doi = "10.1007/JHEP12(2024)215",
    journal = "JHEP",
    volume = "12",
    pages = "215",
    year = "2024"
}

@article{Badger:2021nhg,
    author = "Badger, Simon and Hartanto, Heribertus Bayu and Zoia, Simone",
    title = "{Two-Loop QCD Corrections to Wbb{\textasciimacron} Production at Hadron Colliders}",
    eprint = "2102.02516",
    archivePrefix = "arXiv",
    primaryClass = "hep-ph",
    reportNumber = "CAVENDISH-HEP-21/01",
    doi = "10.1103/PhysRevLett.127.012001",
    journal = "Phys. Rev. Lett.",
    volume = "127",
    number = "1",
    pages = "012001",
    year = "2021"
}

@article{Grazzini:2017mhc,
    author = "Grazzini, Massimiliano and Kallweit, Stefan and Wiesemann, Marius",
    title = "{Fully differential NNLO computations with MATRIX}",
    eprint = "1711.06631",
    archivePrefix = "arXiv",
    primaryClass = "hep-ph",
    reportNumber = "ZU-TH-30-17, CERN-TH-2017-232, ZU-TH 30/17",
    doi = "10.1140/epjc/s10052-018-5771-7",
    journal = "Eur. Phys. J. C",
    volume = "78",
    number = "7",
    pages = "537",
    year = "2018"
}

@article{ATLAS:2022vkf,
    author = "Aad, Georges and others",
    collaboration = "ATLAS",
    title = "{A detailed map of Higgs boson interactions by the ATLAS experiment ten years after the discovery}",
    eprint = "2207.00092",
    archivePrefix = "arXiv",
    primaryClass = "hep-ex",
    reportNumber = "CERN-EP-2022-057",
    doi = "10.1038/s41586-022-04893-w",
    journal = "Nature",
    volume = "607",
    number = "7917",
    pages = "52--59",
    year = "2022",
    note = "[Erratum: Nature 612, E24 (2022)]"
}

@article{CMS:2022dwd,
    author = "Tumasyan, Armen and others",
    collaboration = "CMS",
    title = "{A portrait of the Higgs boson by the CMS experiment ten years after the discovery.}",
    eprint = "2207.00043",
    archivePrefix = "arXiv",
    primaryClass = "hep-ex",
    reportNumber = "CMS-HIG-22-001, CERN-EP-2022-039",
    doi = "10.1038/s41586-022-04892-x",
    journal = "Nature",
    volume = "607",
    number = "7917",
    pages = "60--68",
    year = "2022",
    note = "[Erratum: Nature 623, (2023)]"
}

@article{Huss:2025nlt,
    author = {Huss, Alexander and Huston, Joey and Jones, Stephen and Pellen, Mathieu and R{\"o}ntsch, Raoul},
    title = "{Les Houches 2023 {\textendash} physics at TeV colliders: Report on the Standard Model precision wishlist}",
    eprint = "2504.06689",
    archivePrefix = "arXiv",
    primaryClass = "hep-ph",
    reportNumber = "CERN-TH-2025-071, FR-PHENO-2025-005, IPPP/25/19, TIF-UNIMI-2025-9",
    doi = "10.21468/SciPostPhysCommRep.21",
    month = "4",
    year = "2025"
}

@article{Cepeda:2019klc,
    author = "Cepeda, M. and others",
    editor = "Dainese, Andrea and Mangano, Michelangelo and Meyer, Andreas B. and Nisati, Aleandro and Salam, Gavin and Vesterinen, Mika Anton",
    title = "{Report from Working Group 2}: {Higgs Physics at the HL-LHC and HE-LHC}",
    eprint = "1902.00134",
    archivePrefix = "arXiv",
    primaryClass = "hep-ph",
    reportNumber = "CERN-LPCC-2018-04",
    doi = "10.23731/CYRM-2019-007.221",
    journal = "CERN Yellow Rep. Monogr.",
    volume = "7",
    pages = "221--584",
    year = "2019"
}

@article{Goncharov:2005sla,
    author = "Goncharov, A. B.",
    title = "{Galois symmetries of fundamental groupoids and noncommutative geometry}",
    eprint = "math/0208144",
    archivePrefix = "arXiv",
    doi = "10.1215/S0012-7094-04-12822-2",
    journal = "Duke Math. J.",
    volume = "128",
    pages = "209",
    year = "2005"
}

@article{Goncharov:2010jf,
    author = "Goncharov, Alexander B. and Spradlin, Marcus and Vergu, C. and Volovich, Anastasia",
    title = "{Classical Polylogarithms for Amplitudes and Wilson Loops}",
    eprint = "1006.5703",
    archivePrefix = "arXiv",
    primaryClass = "hep-th",
    reportNumber = "BROWN-HET-1602",
    doi = "10.1103/PhysRevLett.105.151605",
    journal = "Phys. Rev. Lett.",
    volume = "105",
    pages = "151605",
    year = "2010"
}

@article{Duhr:2012fh,
    author = "Duhr, Claude",
    title = "{Hopf algebras, coproducts and symbols: an application to Higgs boson amplitudes}",
    eprint = "1203.0454",
    archivePrefix = "arXiv",
    primaryClass = "hep-ph",
    doi = "10.1007/JHEP08(2012)043",
    journal = "JHEP",
    volume = "08",
    pages = "043",
    year = "2012"
}

@article{Bonciani:2016qxi,
    author = "Bonciani, Roberto and Del Duca, Vittorio and Frellesvig, Hjalte and Henn, Johannes M. and Moriello, Francesco and Smirnov, Vladimir A.",
    title = "{Two-loop planar master integrals for Higgs$\to 3$ partons with full heavy-quark mass dependence}",
    eprint = "1609.06685",
    archivePrefix = "arXiv",
    primaryClass = "hep-ph",
    doi = "10.1007/JHEP12(2016)096",
    journal = "JHEP",
    volume = "12",
    pages = "096",
    year = "2016"
}

@article{Bonciani:2019jyb,
    author = "Bonciani, R. and Del Duca, V. and Frellesvig, H. and Henn, J. M. and Hidding, M. and Maestri, L. and Moriello, F. and Salvatori, G. and Smirnov, V. A.",
    title = "{Evaluating a family of two-loop non-planar master integrals for Higgs + jet production with full heavy-quark mass dependence}",
    eprint = "1907.13156",
    archivePrefix = "arXiv",
    primaryClass = "hep-ph",
    doi = "10.1007/JHEP01(2020)132",
    journal = "JHEP",
    volume = "01",
    pages = "132",
    year = "2020"
}

@article{vonManteuffel:2017myy,
    author = "von Manteuffel, Andreas and Schabinger, Robert M.",
    title = "{Numerical Multi-Loop Calculations via Finite Integrals and One-Mass EW-QCD Drell-Yan Master Integrals}",
    eprint = "1701.06583",
    archivePrefix = "arXiv",
    primaryClass = "hep-ph",
    reportNumber = "MSUHEP-17-002, MITP-16-101, TCDMATH-17-03",
    doi = "10.1007/JHEP04(2017)129",
    journal = "JHEP",
    volume = "04",
    pages = "129",
    year = "2017"
}

@article{Heller:2019gkq,
    author = "Heller, Matthias and von Manteuffel, Andreas and Schabinger, Robert M.",
    title = "{Multiple polylogarithms with algebraic arguments and the two-loop EW-QCD Drell-Yan master integrals}",
    eprint = "1907.00491",
    archivePrefix = "arXiv",
    primaryClass = "hep-th",
    reportNumber = "MITP/19-043, MSUHEP-19-012",
    doi = "10.1103/PhysRevD.102.016025",
    journal = "Phys. Rev. D",
    volume = "102",
    number = "1",
    pages = "016025",
    year = "2020"
}

@article{Chen:2026bxb,
    author = "Chen, Xiang and Grazzini, Massimiliano and Kallweit, Stefan",
    title = "{Two-loop QCD contribution to the quark-initiated production of a top-antitop quark pair and a Higgs boson at the LHC}",
    eprint = "2610.40266",
    archivePrefix = "arXiv",
    primaryClass = "hep-ph",
    reportNumber = "ZU-TH 44/26",
    month = "10",
    year = "2026"
}

@article{Takahasi:1974,
  author  = {Takahasi, Hidetosi and Mori, Masatake},
  title   = {Double Exponential Formulas for Numerical Integration},
  journal = {Publ. Res. Inst. Math. Sci.},
  volume  = {9},
  number  = {3},
  pages   = {721--741},
  year    = {1974}
}

@misc{BoostMath,
  author       = {{Boost.Math contributors}},
  title        = {{Boost.Math}},
  howpublished = {\url{https://www.boost.org/libs/math}},
  note         = {Accessed November 2026}
}

@article{Kerner:1966,
  author  = {Kerner, Immo O.},
  title   = {Ein Gesamtschrittverfahren zur Berechnung der Nullstellen von Polynomen},
  journal = {Numer. Math.},
  volume  = {8},
  pages   = {290--294},
  year    = {1966}
}

@article{Bera:2026qxx,
    author = "Bera, Souvik and Brancaccio, Colomba and Canko, Dhimiter and Ripani, Federico and Zoia, Simone",
    title = "{Two-loop integrals for $t\bar{t}\gamma$ production at hadron colliders in the leading-colour approximation}",
    eprint = "2610.42062",
    archivePrefix = "arXiv",
    primaryClass = "hep-ph",
    reportNumber = "MPP-2026-197, TIF-UNIMI-2026-14",
    month = "10",
    year = "2026"
}
%--------------------------------------------------------------------------------
\end{document}